# A Cancer Biotherapy Resource*

**Preety Priya[1] and Vicente M. Reyes, Ph.D.[2], [3]**

(*, M. S. Thesis; [1], M. S. student; [2], thesis advisor)
[3], E-mail: **vmrsbi.RIT.biology@gmail.com**

Submitted in partial fulfillment of the requirements for the **Master of Science** degree in Bioinformatics at the Rochester Institute of Technology

**Preety Priya**

Dept. of Biological Sciences, School of Life Sciences
Rochester Institute of Technology
One Lomb Memorial Drive, Rochester, NY 14623

August 2012

8-13-2012

# Cancer biotherapy resource

Preety Priya

# Cancer Biotherapy Resource

*Preety Priya*

*Masters of Science in Bioinformatics*

**Approval Committee:**

Dr. Irene Evans, Professor, RIT

Dr. Jim Leone, Professor, RIT

Dr. Michael V. Osier, PhD., Associate Professor, RIT

*8/13/2012*
*Thomas H. Gosnell School of Life Sciences*
*College of Science, Bioinformatics Program*
*Rochester Institute of Technology*

*This thesis is dedicated to my loving family, to my parents who supported and encouraged me throughout my life and my brother who has always been helpful and understanding.*

# DISSERTATION AUTHOR PERMISSION STATEMENT

**TITLE OF THESIS:** Cancer Biotherapy Resource

Author: Preety Priya

Degree: Masters

Program: Bioinformatics

College: College of Science, Rochester Institute of Technology

I, Preety Priya, understand that I must submit a printed copy of my thesis or dissertation to the RIT archives, per current RIT guidelines for the completion of my degree. I hereby grant to the Rochester Institute of Technology and its agents, the non-exclusive license to archive and make accessible my thesis or dissertation in whole or in part in all forms of media in perpetuity. I retain all other ownership rights to the copyright of the thesis dissertation. I also retain the right to use in future works (such as articles or books) all or part of this thesis or dissertation.

\_\_\_\_\_\_\_\_\_\_\_\_\_\_\_\_\_\_

Preety Priya

(Date)

# ACKNOWLEDGEMENTS

I feel immensely happy and privileged to finish my Masters of Science degree from the Rochester Institute of Technology. Coming to the United States was the first time I had been outside my home country India, and RIT made my stay a memorable experience. The thesis marks an end to the magnificent two years I spent here, and many of the people I encountered made the time more comfortable and certainly memorable.

A special thanks to Dr. Vicente Reyes and Dr. Jim Leone from the Rochester Institute of Technology for their invaluable guidance and support. I would like to express my sincere thanks to Dr. Gary Skuse, Dr. Anne Haake, Dr. Irene Evans and Dr. Michael Osier for their timely help and support throughout my Masters.

I would also like to thank Nicoletta Bruno Collins for her help with the academic formalities. I would like to thank my Professors, faculty, family and friends for their assistance.

Last, but not the least, I would like to thank the faculty from the International Students Services Office who were a second family for me and many other Internationals.

# Abstract

'Cancer Biotherapy' – as opposed to cancer chemotherapy- is the use of macromolecular, biological agents instead of organic chemicals or drugs to treat cancer. Biotherapy is a treatment modality that blocks the growth of cancer cells by interfering with specific, targeted molecules needed for carcinogenesis and tumor growth instead of simply interfering with rapidly dividing cells as in chemotherapy[1]. In light to the much higher selectivity of biological agents than chemical agents for cancer cells over normal cells, there is a much less toxic side effect in biotherapy as compared to chemotherapy. As solid tumor cancer continues to be analyzed as a chronic condition, there is an absolute need for long-term treatment with minimal side effects. The International Society for Biological Therapy of Cancer, being the only available information database for cancer biotherapy, lacks some crucial information about various cancer biotherapy regimens and the information presented seemed unorganized and unsystematic making it difficult to search for results. With the increasing rate of cancer deaths across the world and biotherapy studies, it is acutely necessary to have a comprehensive curetted cancer biotherapy database. The database accessible to cancer patients and also should be a sounding board for scientific ideas by cancer researchers.

The database/web server has information about main families of cancer biotherapy regimens to date, namely, 1.) Protein Kinase Inhibitors, 2.) Ras Pathway Inhibitors, 3.) Cell-Cycle Active Agents, 4.) MAbs (monoclonal antibodies), 5.) ADEPT (Antibody-Directed Enzyme Pro-Drug Therapy), 6.) Cytokines (interferons, interleukins, etc.), 7.) Anti-Angiogenesis Agents, 8.) Cancer Vaccines (peptides, proteins, DNA), 9.) Cell-based Immunotherapeutics, 10.) Gene Therapy, 11.) Hematopoietic Growth Factors, and 12.) Retinoids 13.) CAAT. For each biotherapy regimen, we will extract the following attributes in populating the database: (a.) Cancer type, (b.) Gene/s and gene product/s involved, (c.) Gene sequence (GenBank ID), (d.) Organs affected (e.) Chemo treatment, (f.) Reference papers, (g.) Clinical phase/stage, (h.) Survival rate (chemo. Vs. biother.), (i.) Clinical test center locations, (j.) Cost, (k.) Patient blog, (l.) Researcher blog, (m.) Future work.

The database accessible to public through a website and had FAQs for making it understandable to the laymen and discussion page for researchers to express their views and ideas. In addition to information about the biotherapy regimens, the website is linked to other biologically significant databases like structural proteomics, metabolomics, glycomics, and lipidomics web servers. Also, the websites presented the news in the field of biotherapy and other links which are relevant from biotherapy point of view. The database attributes would be regularly updated for novel attributes as discoveries would be made.

# Table of Contents

# Figure Contents

# Introduction

Cancer is not just one ailment it is a saga of diseases. The versatility and adaptability of the cancer cells makes them difficult for scientists to develop methods for curing this disease. While some cancers remain tucked inside darker recess of the human body like liver or colon cancer where they are extremely difficult to spot and treat, some others like melanoma or retinoblastoma are near to the surface and easily observable and treated. Cancer cells also carry the characteristic of growing remarkably fast, like in case of leukemia, and exceedingly slow, as in follicular lymphoma.

Medically termed as malignant neoplasm, cancer is an uncontrolled growth of abnormal cells in the body. This uncontrolled growth of cells can be attributed to mutations in the signals that regulate the cell cycle of growth and division.

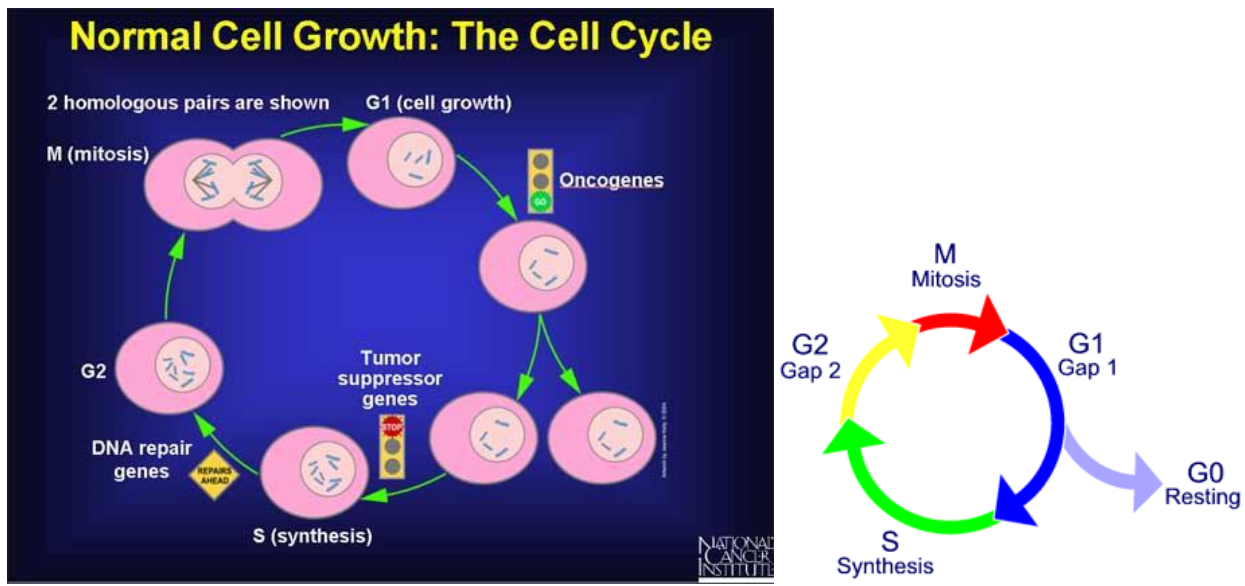

**Fig 1: a. The cell cycle; b. Cell cycle phase**

[2]Most cells remain in interphase, the period between cell divisions, during the cell cycle. The initial stage of the interphase, G1 (for first gap) involving rapid growth, metabolic activity and

synthesis of RNA. Next is S phase (for DNA synthesis), where cell continues to grow, and DNA gets replicated. In G2 (for second gap), the cell continues to grow and prepares for cell division. Next is Cell division (mitosis) called as M phase. Cells that do not divide for long periods do not replicate their DNA and are in G0. In normal cells, tumor suppressor genes act as braking signals during G1 to stop or slow the cell cycle before S phase. DNA repair genes are active throughout the cell cycle, particularly during G2 after DNA replication and before the chromosomes prepare for mitosis.

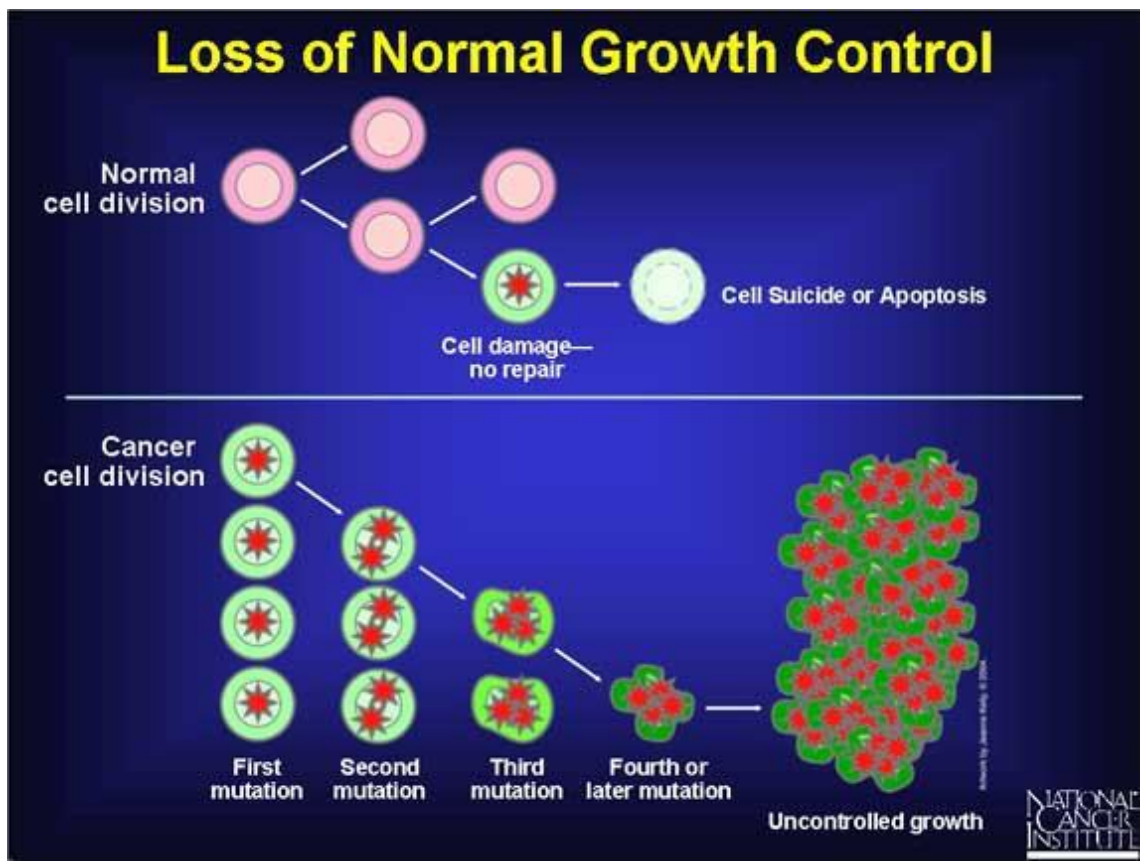

**Fig 2: Cell division: normal cell vs. cancerous cell**

Cancer cells overcome the normal cell signaling process and become immortal. Apart from cell proliferation becoming uncontrollable, cells becoming resistant to apoptosis (programmed cell death) is another reason behind cancerous cells. In regards to mutations in tumor suppressors or activation of proto-oncogenes to oncogenes, the cell cycle gets disrupted, and the cells divide uncontrollably. Normal cells remain in the area where they belong and do not spread to other parts of the body. Cancer cells disregard this principle and spread through the body in a process called metastasis. These include direct invasion and destruction of the organ of origin, or spread through the lymphatic system or bloodstream to distant organs. The immune system consists of a group of cells called white blood cells that recognize and destroy "foreign" material in the body such as bacteria, viruses, and unfamiliar or abnormal cells. Cancer cells manage to slip through this detection system without triggering the immune system to start fighting, either at the primary cancer site, in the blood vessels, or at the site of the distant spread[3].

Although early detection coupled with early intervention is the best way to fight cancer, an improved modality over current chemotherapeutic regimens for treating cancer is evidently required. Presently available chemotherapeutic agents lack the level of selectivity required for a distinction between normal cells and cancer cells and as a result, produce serious, sometimes fatal side effects. A significant proportion of those who die from cancer die from the ill effects of the chemotherapeutic drugs. Biotherapeutic agents employ large polymeric macromolecules on contrast to the chemotherapeutic agents which are mostly small organic ligands. With the makeable increase in the deaths due to cancer across the country and biotherapeutic studies getting proved beneficial, over chemotherapy there is a profound need of a database of cancer biotherapy studies which is curetted and comprehensive for use as a resource to researchers and patients alike.

## Chemotherapy

Chemotherapy is the treatment of cancer with an antineoplastic drug or a combination of such drugs into a standardized treatment regimen[4].

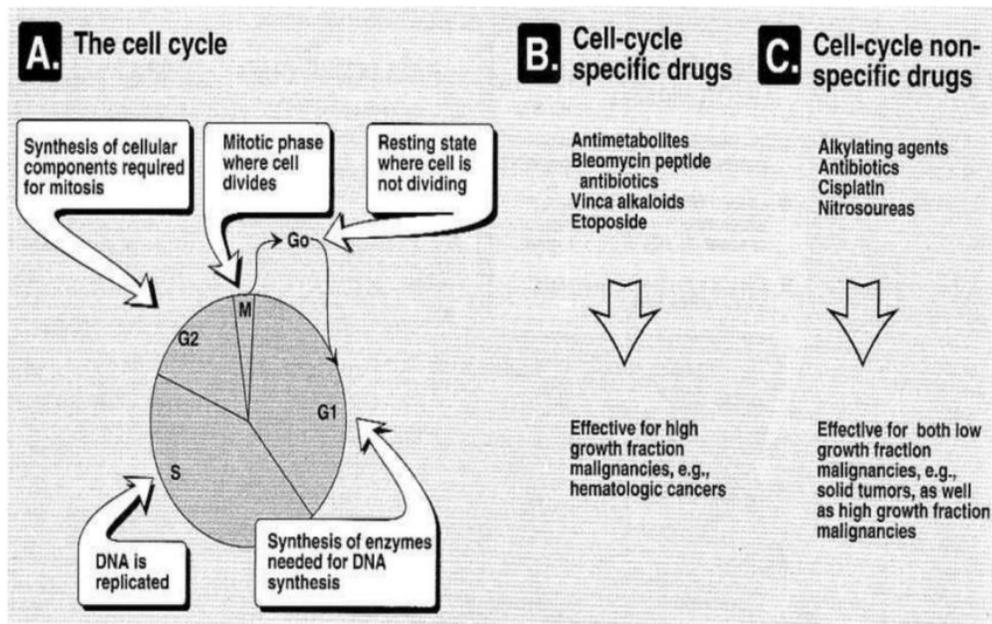

Figure 38.4
Effects of chemotherapeutic agents on the growth cycle of mammalian cells.

**Fig 3: Cancer Chemotherapy**

Chemotherapy is designed to kill cancer cells and is administered through veins, injected into a body cavity delivered orally in the form of a pill, depending on which drug is used[5]. It works by destroying fast growing cells but cannot distinguish cancer cells from the normal ones so is the less specific treatment. As a result, normal fast growing body cells also gets destroyed like hair, blood and gastrointestinal cells.

Cancer cells differ in growing and spreading some grow fast in leukemia while some grow unusually slow like in follicular lymphoma. This is the reason, different types of chemotherapy drugs target the growth patterns of different types of cancer cells. Each drug has a different way of working and is effective at a specific time in the life cycle of the cell it targets.

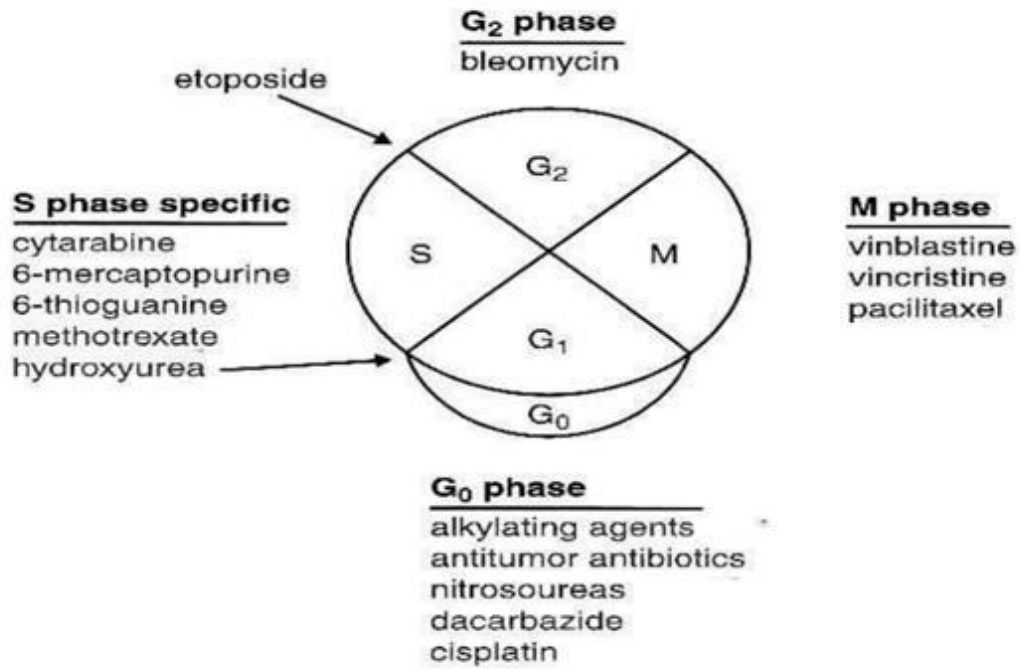

**Fig 4: Chemotherapeutic drugs in cell cycle**

Chemotherapeutic drugs can be divided in to alkylating agents, antimetabolites, anthracyclines, plant alkaloids, topoisomerase inhibitors, and other antitumor agents. All the drugs eventually affect the cell division and DNA synthesis. Alkylating agents like cisplatin and carboplatin impair cell function by forming covalent bonds with the amino, carboxyl, sulfhydryl, and phosphate groups in biologically relevant molecules. They modify the DNA chemically. Drugs like mercaptopurine and azathioprine are antimetabolite as they prevent purines and pyrimidines from becoming incorporated in to DNA during the "S" phase of the cell cycle, stopping normal development and division. They also affect RNA synthesis and due to their efficiency these drugs are the most widely used cytostatics. Plant alkaloids like vinca alkaloids (vinblastine, vincristine etc.): Inhibits the assembly of tubulin into microtubules during M phase, podophyllotoxin (etoposide and teniposide): prevents the cell from entering the G1 phase and the replication of DNA in the S phase and taxanes (docetaxel): enhance stability of microtubules, preventing the separation of chromosomes during anaphase. Topoisomerase inhibitors include irinotecan and topotecan (type I topoisomerase inhibitors), amasacrine, etoposide (type II topoisomerase inhibitors). They inhibit type I and type II topoisomerase and interferes with the transcription and replication of DNA by disrupting DNA super coiling.

Dosage of chemotherapy is confusing as a too low dose can be ineffective against tumor and excessive dose can prove intolerable to patient. Chemotherapeutic techniques have a range of severe side-effects also which includes weakening of the immune system resulting in fatal infection, blood loss leading to severe anemia, infertility, heart damage and most importantly, development of secondary neoplasm.

With all these adverse effects during and after the treatment, cost of treatment being expensive without any guarantee of curing the disease, there is an absolute need of exploring the area of biotherapy in the treatment of cancer.

**Biotherapy**

Biotherapy is the treatment technique that uses the body's own immune system to fight cancer. Biotherapy can be used to halt or suppress the processes that allow cancer growth, help the immune system identify cancer cells, and promote the body's natural ability to repair or replace cells damaged by other cancer treatments. Biotherapy can fight cancer by slowing down the cancer cell growth, making it easier for immune system to destroy cancerous cell and keeping cancer from spreading to other parts of the body.

**Table 1: Biotherapy Vs Chemotherapy**

| **Cancer Chemotherapy** | **Cancer Biotherapy** |
|---|---|
| • Involves small organic molecules (monomer) | • Involves large macromolecules (polymer) |
| • Large scale production is done by organic chemical synthesis | • Large scale production involves cloning methods |
| • Commercial production is less expensive | • Very expensive commercial production |
| • Prone to have side effects | • Much less chance of toxic side effects and palliates as the treatment stops |
| • Cancer cells may develop resistance to the chemotherapeutic agents | • Development of resistance to biotherapeutic agent is less likely |
| • High dose and systemic modality required to counteract low selectivity and maximize anti-tumor activity | • Due to high selectivity, only low does are usually required |

Biotherapy does not attack healthy cells, but it can still have some side effects. The most common are skin rashes and flulike symptoms -- body aches, chills and fever. There are some side effects of biotherapy also which palliates as the treatment gets completed. A major drawback of biotherapy is the high cost of treatment which restricts it to be available for everyone.

# Biotherapy Regimens

***Cytokines:***

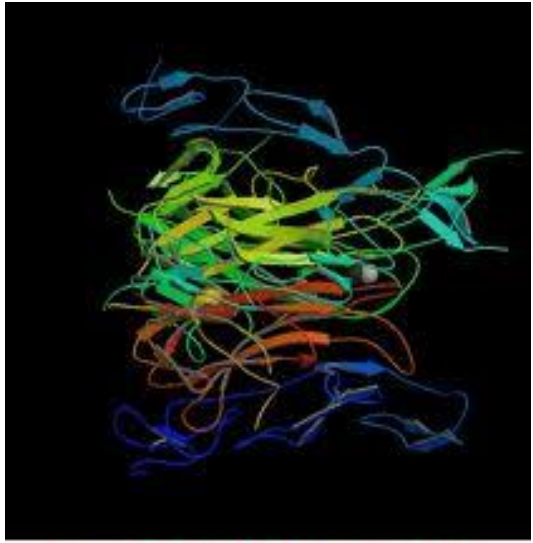

Interaction of TNFSF10 (cytokine) and TNFRSF10B (cytokine-receptor) PDB ID:1D0G   ***Fig.5: Cytokine crystal***

Cytokines are small cell signaling protein molecules secreted by innate and adaptive immune systems and are responsible for carrying cell signals to maintain intercellular communication. The term "cytokine" encompasses a large and diverse family of regulators produced throughout the body by cells of diverse embryological origin[6]. Cytokines are immunomodulating agents often confused with hormones between biologists. The cytokines have the systemic immunomodulating effect which differentiates it from hormones as their activity and effects are local. Also, cytokines actions can be autocrine and paracrine but not endocrine like hormones. Every cytokine has a matching cell-surface receptor which ultimately activates the intracellular signaling cascade to regulate cell functioning. This may include the up regulation and/or down regulation of several genes and their transcription factors, resulting in the production of other cytokines, an increase in the number of surface receptors for other molecules, or the suppression of their own effect by feedback inhibition[6]. Deregulations in cytokine activities have led to many adverse effects like Alzheimer's disease, cancer etc. Of all the cytokines identified so far, interleukins (IL -2 & 12), interferons (IFN -α) and granulocyte-macrophage colony stimulating factor (GM-CSF) recognized as vital contributors in cancer research and treatment.

Interferon was isolated in 1957 by Isaacs and Lindenmann which interferes with the viral activity in cells. IFN family comprises of a remarkably complex set of protein and glycoprotein's heterogenous in nature.

They can be classified as:

Type I: IFN –α, -β, -τ and –ω

Type II: IFN –γ

Type I IFNs has a different structure and interaction with cell surface receptors from type II. Also, type I are more effective in inducing antiviral activity in the cell than type II. All the type I IFNs share a common ligand binding site and induce common biological effects. IFN –α and –ω is derived from leukocytes, IFN -β is derived from fibroblasts and IFN –τ comes from trophoblasts. IFN –γ is secreted by CD8+ T cells and some CD4+ T cells when activated by interleukins. As an anticancer agent, IFN – α received its first FDA approval in 1986, for its treatment of hairy cell leukemia. IFA – α is also involved in the studies of melanoma, non-Hodgkin's Lymphoma and chronic myelogenous leukemia. IFN –α activates several pathways, which, can induce, inhibit or modify the effects of several cytokines like IFN-γ, IL-1, IL-2, IL-6, IL-8 and TNF- α. Exposure of NK cells to IFN- α significantly up regulates the activity of NK cells through accelerated kinase, IL-2 dependent growth, mediation of antibody-dependent cellular cytotoxicity. Type I IFNs also regulates the activation and function of macrophages and monocytes, which, results in an increase in tumoricidal activity of macrophages. Immunomodulation of macrophages by IFNs show alteration in its characteristics and increases the expressions of receptors for Fc portion of immunoglobulin. Finally, the up regulation of Fc receptor promotes increased phagocytosis by macrophages thereby enhancing the lytic activity[7].

Interleukins is another area of biotherapy which is intensively studied. They are a group of cytokines expressed by white blood cells. Interleukins stimulates the growth of blood cells and regulate inflammatory and immune responses. IL-2 has found an important role in effective biotherapy studies of melanoma, lymphoma and renal cancer. It regulates the growth of T cells in the thymus and enhances the function of T cells and NK cells. IL-2 also activates the lymphoma activated killer cells (LAK). These LAK cells are killer T cells and destroy tumor cells.

### *Colony Stimulating Factors:*

Colony-stimulating factors are substances that stimulate the production of blood cells and promote their ability to function. They do not directly affect tumors, but through their role in stimulating blood cells they can be helpful as support of the person's immune system during cancer treatment. CSFs is playing an pivotal role in cell proliferation, suppression of apoptosis, cell differentiation, induction of maturation and functional stimulation of mature cells. They also play important role involved in stem cell transplant. They are also known as biological response modifiers in cancer biotherapy studies.

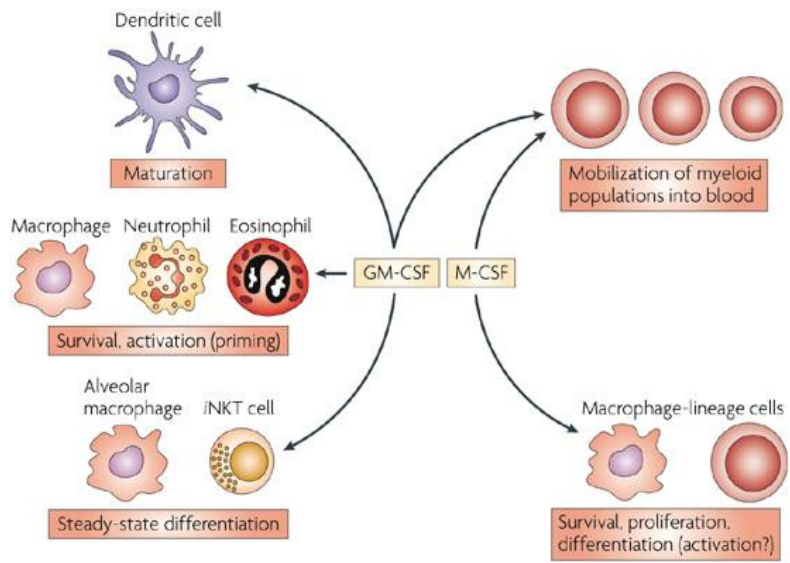

**Fig.6: Colony Stimulating Factor**

HGFs is regulators of blood cell proliferation and development. They augment hematopoiesis during bone marrow dysfunction. Hematopoietic Growth Factors are a group of substances with the ability to support hematopoietic (blood cell) colony formation in vitro. This group of substances includes erythropoietin, interleukin-3 and colony-stimulating factors (CSFs). Erythropoietin stimulates production of erythrocytes, or red blood cells. Interleukin-3 and CSFs can mature cells, have overlapping capabilities to affect progenitor cells ("parent" cells that will develop into a specific type of cell) of some blood cell lines and can also affect cells outside the hematopoietic system. HGFs is used to promote bone marrow proliferation in aplastic anemia, following cytotoxic chemotherapy, or following a bone marrow transplant. The American Society of Clinical Oncology has made guidelines for the use of CSFs after chemotherapy. GM-CSF gene-modified tumor vaccines have been tested in cancer patients. Some phase I and phase II studies have been performed that showed induction of immunity without significant toxicity.

*Retinoids:*

Retionoids are a group of natural and synthetic analogs of vitamin A. They are significant for vision, growth, reproduction and immune function. In the biotherapy research, retinoid is involved in arresting or reversing the process of carcinogenesis to prevent cancer invasion and metastasis and treatment of cancer (Singh and Lippman, 1998). The significant results in the treatment of acute promyelocytic leukemia (APL), metastatic squamous cell cancer of skin and cervix drew attention of specialists to obtain investigational new drug application (INDA) status for the therapeutic use of retinoid (Smith et al., 1992). Retinoids used in clinical trials are all-trans-retinoic acid, 4-hydroxyphenyl all-trans-retinoic acid amide 13 cis-retinoic acid and 9-cis-retinoic acid.

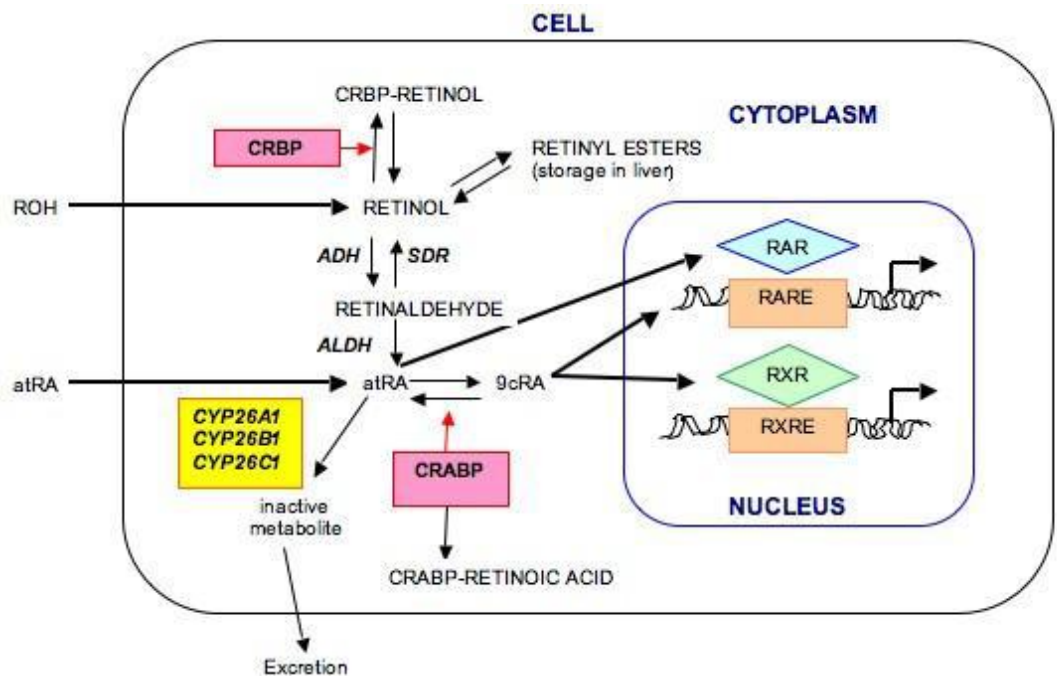

**Fig 7: Retinoid metabolism and mechanism of action**

[8]They play a crucial role in cell differentiation and can suppress tumor promotion. They modify some properties of fully transformed malignant cells by activating or repressing

specific genes. Retinoids receptors are expressed in normal and malignant epithelial breast cells. Binding of retinoid to the nuclear receptors (i.e. retinoic acid receptor (RAR)-α, -β and -γ and retinoid X receptor (RXR)-α, -β and -γ), which are ligand-activated transcription factors, leads to regulation of several cellular processes, including growth, differentiation and apoptosis[8].

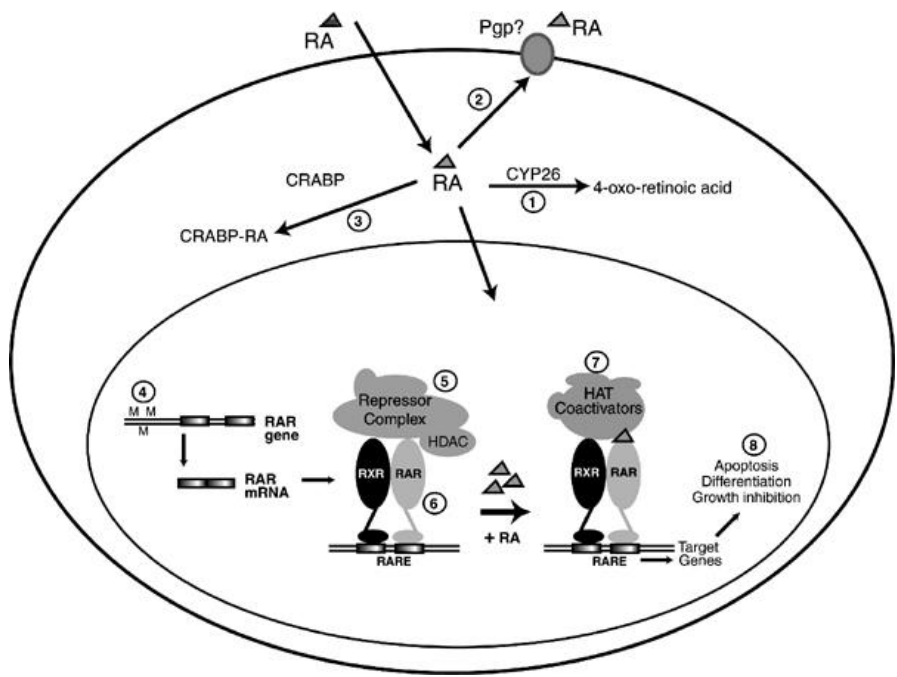

**Fig 8: Potential mechanisms of RA resistance. Cellular retinoid resistance may occur through (1) increased *P*450 catabolism, (2) drug export (P-glycoprotein (Pgp) mediated), (3) sequestration of retinoid by CRABPs or other proteins, (4) decreased expression of RARs through promoter methylation, as depicted (M), (5) persistent histone deacetylation, (6) RAR rearrangement or mutation in the RAR ligand-binding domain, (7) coactivator alteration, or (8) alterations downstream of target gene expression**

*Monoclonal Antibody:*

Antibodies are proteins produced by B lymphocytes of the immune system in response to foreign proteins called antigens. They are carriers to deliver drugs and toxins to tumor cells. Monoclonal antibodies are single clone of cells through hybridoma technology. They have been extensively studied in the investigation and clinical trials of cancer diagnosis and treatment.

[9]For the production of monoclonal antibodies, mouse is immunized against a target cell. This immunization allows the mouse to produce antibodies against that antigen. These antibodies against the specific cell antigen are then isolated from the mouse spleen and fused with the tumor cells grown in culture. The resulting fused cells are called hybridomas. Hybridomas is continuously growing cell lines due to the fusion of myeloma cells with normal cells capable of producing antibodies. Each hybridoma will produce a large quantity of identical antibodies. They are also called monoclonal antibodies as they are produced by the identical offspring of a single cloned antibody producing cell[9].

MABs is used to classify leukemia and lymphoma. When antibody binds to a cell the crystallizable fragment portion of immunoglobulin can be bound by effector cells like monocyte or lymphocyte leading to the destruction of target cancerous cell. This is known as antibody dependent cellular cytotoxicity (ADCC). Another approach can be directing antibodies against growth factor receptors of the cancer cells to check their growth. There are three MABs approved for the cancer treatment: Rituximab for non-Hodgkin's lymphoma, Trastuzumab for metastatic breast cancer and Gemtuzumab for acute myeloid leukemia[9].

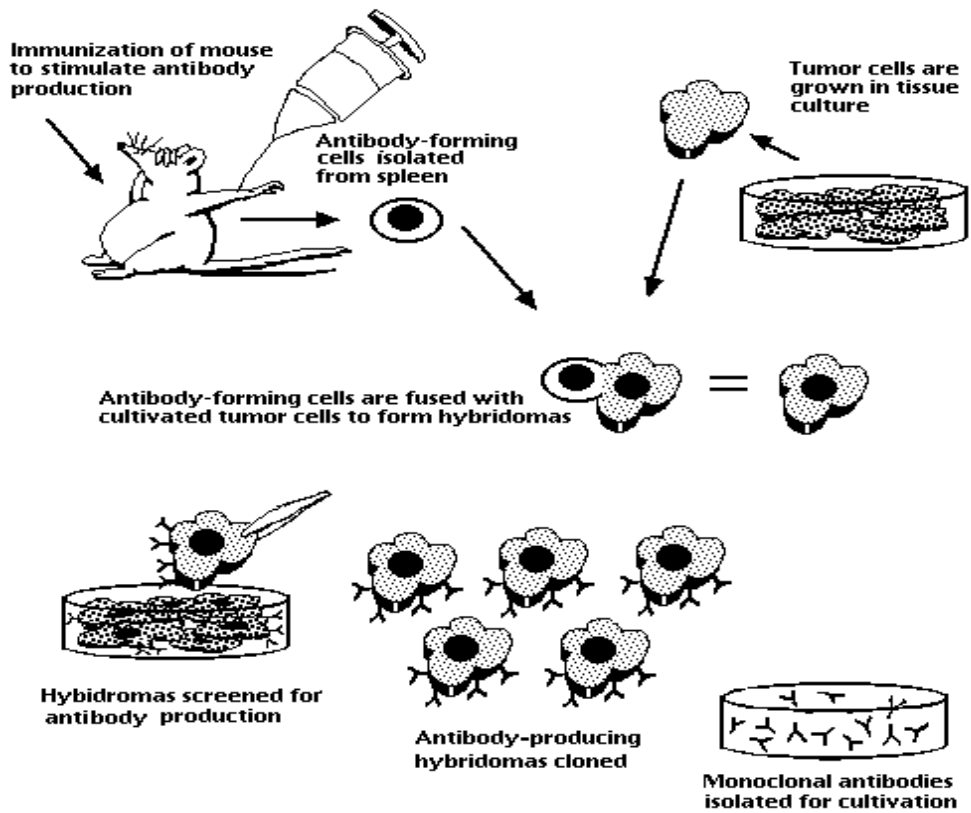

**Fig.9: A summary of the process of monoclonal antibody production**

Monoclonal antibodies have been employed in many ways to study cancer and in cancer therapy including diagnosis, monitoring and treatment of the disease. They aid in diagnosis, such as the application of flow cytometry in the identification of different subsets of non-Hodgkin's lymphoma[10]. Monoclonal antibodies can act against specific antigens on cancer cells and can aid the patient's immune response. Monoclonal antibodies can be programmed to act against cell growth factors, thus blocking cancer cell growth. They can be linked to anticancer drugs, radioisotopes and other biological response modifiers. When the antibodies bind with antigen-bearing cells, they deliver their load of toxin directly to the tumor. Monoclonal antibodies may also be used to preferentially select normal stem cells from bone marrow or blood in preparation for a hematopoietic stem cell transplant in patients with cancer.

Monoclonal antibodies have many ways of attaining their therapeutic effect. They can have a direct effect in programmed cell death or apoptosis. They can block growth factor receptors, effectively arresting proliferation of tumor cells. In cells that express monoclonal antibodies, they can bring about anti-idiotype antibody formation. They also have an indirect effect which involves recruiting cytotoxic cells like monocytes and macrophages and is termed as antibody –mediated cytotoxicity (ADCC). They also bind with complement leading to cytotoxicity known as complement dependent cytotoxicity (CDC) [10].

## *Cancer Vaccines:*

Vaccines are medicines that boost the immune system's natural ability to protect the body against "foreign invaders," mainly infectious agents that may cause disease[11].

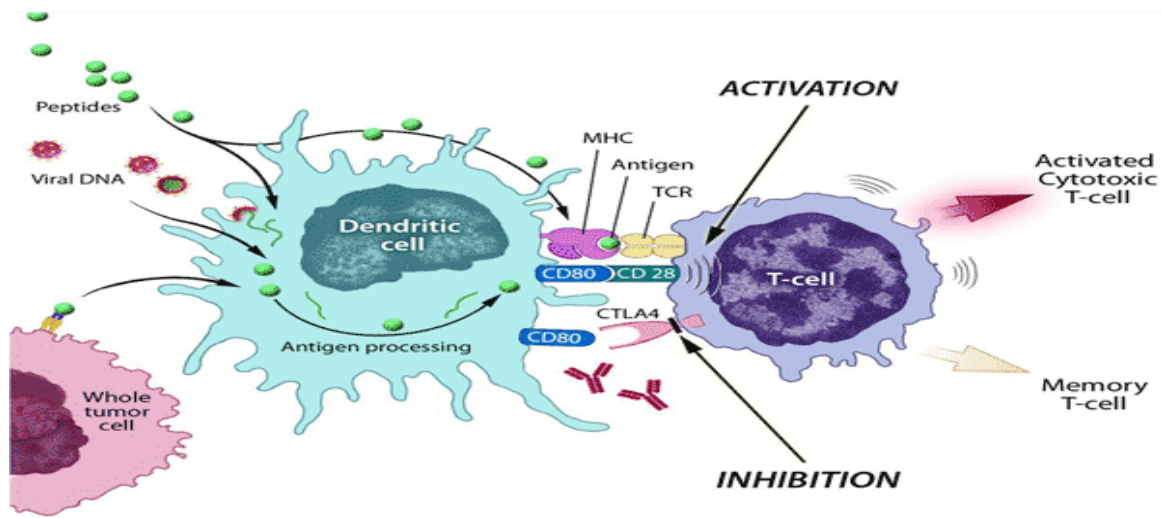

**Fig 10: Cancer Vaccine, FDA Approves First Therapeutic Cancer Vaccine**

[12,13]Vaccines is a form of active immunotherapy as they generate intrinsic immune response. They specifically target the antigen so fall in the category of targeted therapy. Cancer vaccines are made from the person's own cancer cells. The cancer cells are treated with radiotherapy or heat to prevent them from multiplying and making them harmless to be used as vaccines. Cancer vaccines can be made out of antigens from the cancer cell surface or the entire cell sometimes. Cancer vaccines are designed to contain substances that boost the immune system. As the cancer vaccine contains similar proteins to the cancer cells, it is hoped that the immune system will be stimulated to attack and destroy them.

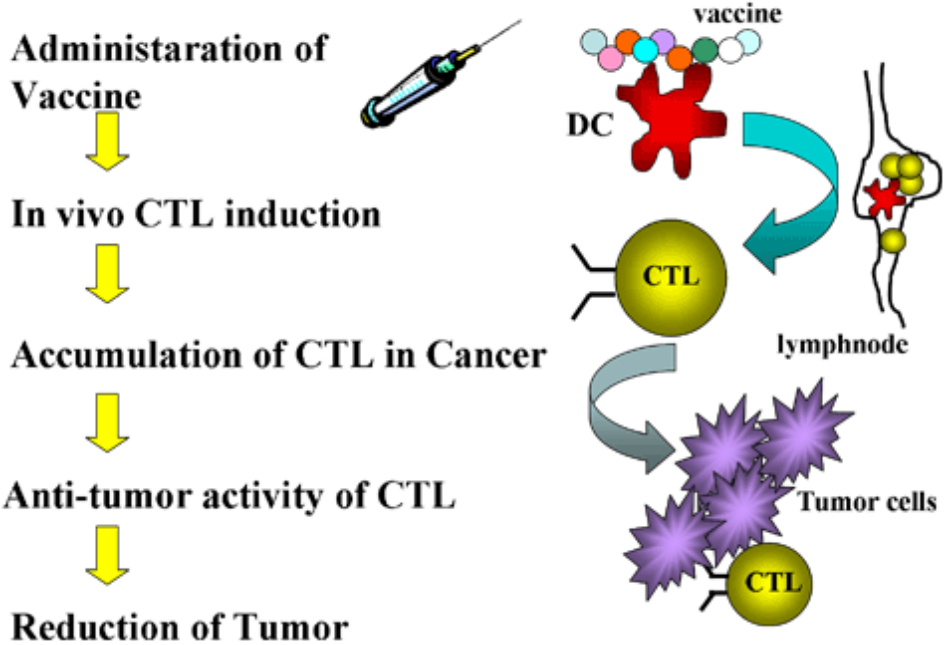

**Fig.11: Antitumor mechanism of Cancer Vaccine,**

Cancer vaccines are designed to stop the cancer cell growth, shrink tumor cells and also to prevent cancer relapses. Cancer cells carry normal self-antigens in addition to specific cancer-associated antigens. Furthermore, cancer cells sometimes undergo genetic changes that may lead to the loss of cancer-associated antigens. Finally, cancer cells can produce chemical messages that suppress anticancer immune responses by killer T cells. As a result, even when the immune system recognizes a growing cancer as a threat, the cancer may still escape a strong attack by the immune system. Cancer vaccines are designed to activate B cells and killer T cells by directing them to recognize and attack cancer cells. They introduce antigen molecules in the body through injection thereby stimulating the immune response.

Patient's tumor cells and blood cells are compared to find the specific point mutations. These mutations are searched for the patient's MHC class I compliments. Once, the complementary sequence is found, antigens with those sequences are injected in the body through cancer vaccines. This antigen administration activates TLRs to recognize the cancer cells exhibiting this antigen and MHC class I binds to the cancer cells leading to their destruction by T cells[12, 13].

*ADEPT (Antibody Directed Enzyme Prodrug Therapy):*

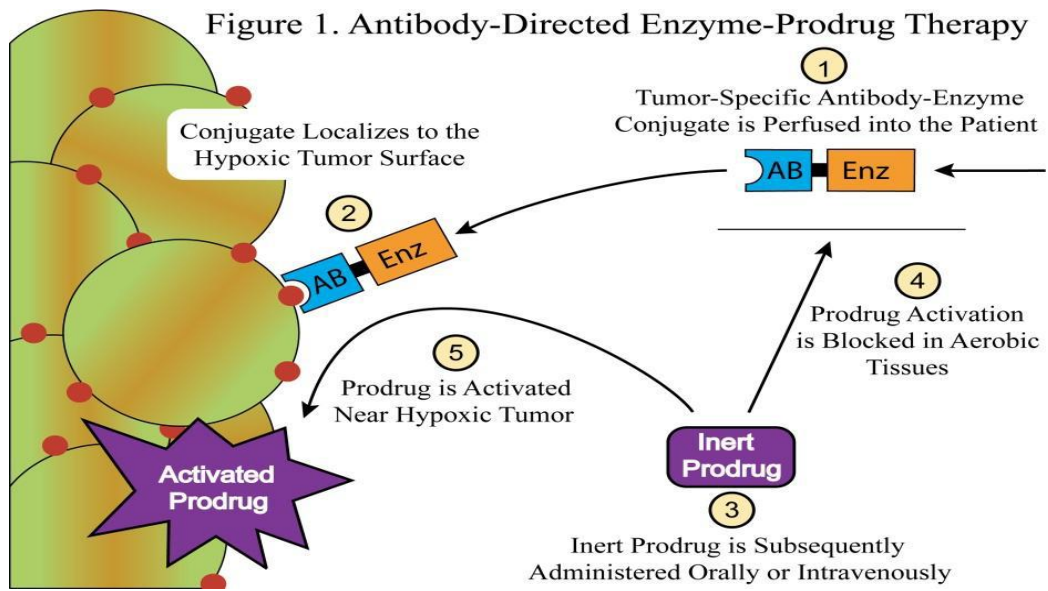

**Fig.12: ADEPT, designed enzyme for the treatment of cancer**

[14]The main objective of ADEPT is to selectively deliver chemotherapy drugs to the cancer affected site in the body. The basic principle of ADEPT is to target enzyme to tumors by attaching it to an antibody directed to a tumor associated antigen. The non-toxic prodrug is given which in the presence of enzyme gets converted to tumor cell killing agent. The main objective of ADEPT is to facilitate the delivery of chemotherapeutics drugs to the tumors accurately. One of the basic concepts involved is that cancer cells may express antigens in high levels that are only expressed in low levels in normal cells. This indicates that antibodies targeted against these antigens will preferentially bind to the cancer cells. Anything stuck to the antibody under these circumstances will be specifically delivered to the cancer cells which forms the basic foundation of the concept of antibody directed enzyme prodrug therapy.

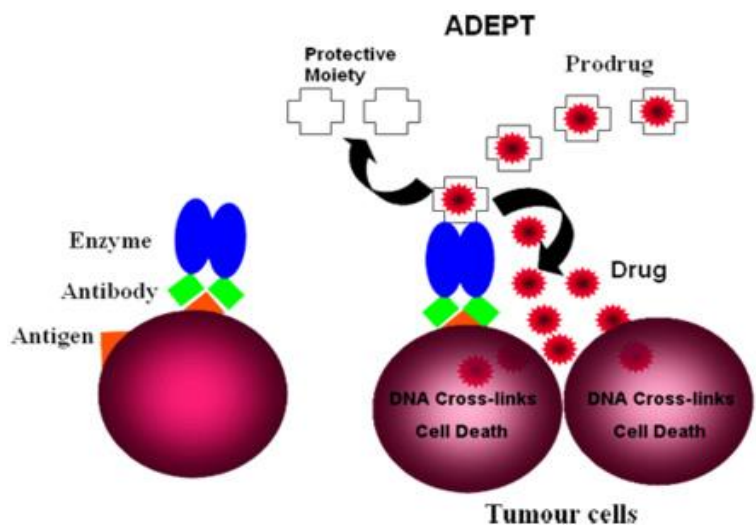

**Fig.13: ADEPT in cancer therapy**

[15]In initial stage of ADEPT, an antigen expressed on tumor cells binds an antibody–enzyme conjugate. A prodrug is then administered and is converted to an active cytotoxin only in the environs of the tumor. The utility of chemotherapeutic drugs is severely hampered, however, by their limited selectivity for tumor cells over normal cells. Enzyme prodrug

therapy (EPT) attempts to circumvent this obstacle by using a nontoxic prodrug which is converted into a cytotoxic product only in the vicinity of the tumor. This strategy exploits the targeting of activating enzyme to the vicinity of the tumor using monoclonal antibodies specific to particular cancer types covalently attached to the activating enzyme[15].

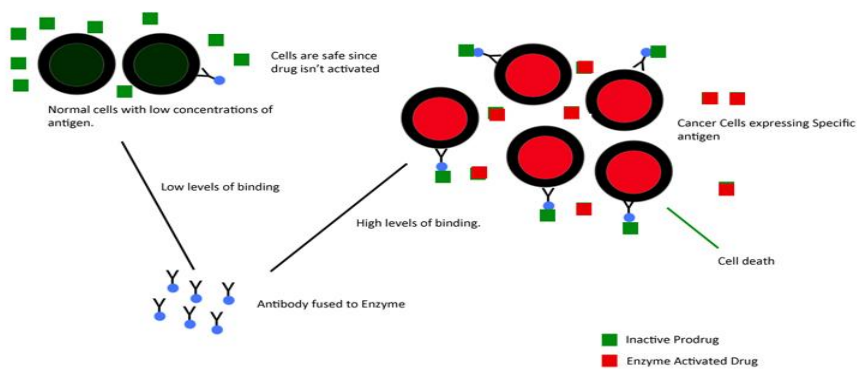

**Fig.14: How ADEPT Works - Antibodies, which bind preferentially to cancer antigens, have enzymes fused to them, which activates a non-toxic prodrug to an active drug only where tumors are present, increasing efficacy and lowering toxicity to normal cells.**

[15]1. Antibodies specific to antigens can be prepared by monoclonal antibody technology.

2. Suitable enzyme is selected to convert the non-toxic prodrug to toxic drug at the site of cancer.

3. The enzyme-MAbs conjugates are then injected into the body where they find themselves concentrating at the tumors.

4. The prodrug is then injected which circulates in the body and being non toxic does not affect other normal cells.

5. When the prodrug comes in the vicinity of the cancer cells which have a high concentration of activating enzymes due to antibody directed binding, it is converted to the active form which is capable of killing tumor cells. This indicates that it is possible to have drugs exerting their effects only at the location of tumors, thereby preventing problems with the toxicity during systemic circulation.

Along with the positive sides the treatment holds certain potential problems. The drug may cause toxicity after activation in tumors until it is excreted. Some of the drug and chemo resistant cancer cells may express genes like MDr1 which can pump drugs out of the cells, this will make ADEPT ineffective[15].

### *Anti-Angiogenesis Agents:* [16, 17]

Angiogenesis is the process of formation of new blood vessels involving migration, growth and differentiation of endothelial cell. The process of angiogenesis is controlled by chemical signals in the body. These signals can stimulate both the repair of damaged blood vessels and the formation of new blood vessels.

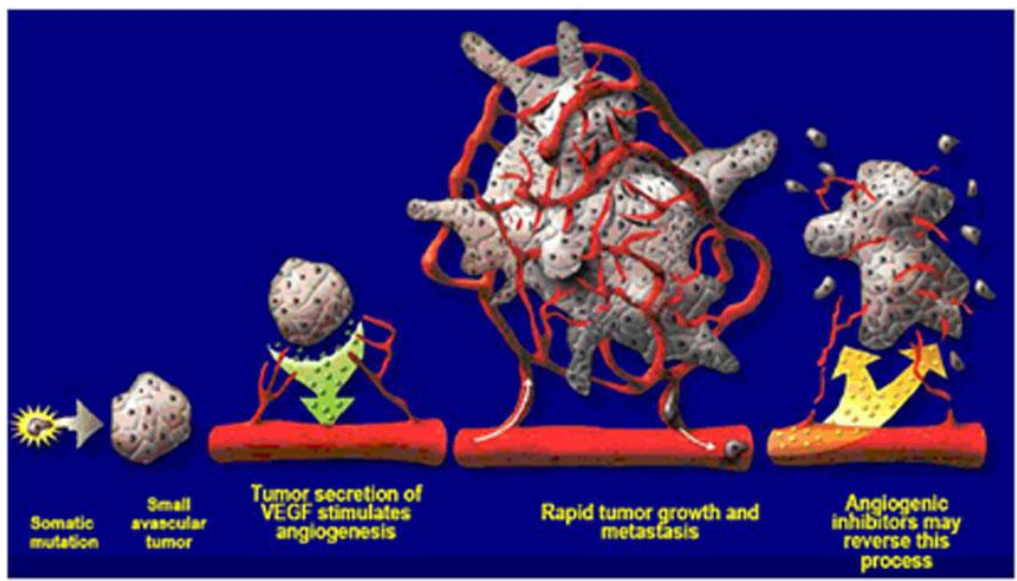

**Fig 15: Anti-Angiogenesis Agents, Understanding Cancer Series, NCI**

In cancer, the process of angiogenesis creates new and ultra small blood vessels that give a tumor its own blood supply and allow it to grow. So, anti- angiogenesis focuses on restriction of the blood supply to the tumor cells by preventing or disrupting the growth of blood vessels supporting tumor cells.

Angiogenesis inhibitors, interfere with blood vessel formation by stimulating and inhibiting effects of chemical signals so that blood vessels are formed when needed. Angiogenesis inhibitors interfere by attaching to VEGF and inactivating VEGF receptor.

### *Cell-based Immunotherapeutics*[18]*:*

In cellular therapy, processed tissue from animal embryos, fetus or organs are injected or taken orally. Tissues are obtained from specific organs in correspondence with the unhealthy organs or tissues of the recipient. The body automatically transports the injected cells to the target organs to strengthen and regenerate the specific organ. The organs and glands used in cell treatment include brain, pituitary, thyroid, adrenals, thymus, liver, kidney, pancreas, spleen, heart, ovary, testis, and parotid. In 1970, Wolfram Kühnau, MD, an associate of Dr. Niehans, began using cellular therapy to treat cancer patients in Tijuana, Mexico.

Cellular therapies involve:

1. Lymphokine-activated killer cells

2. Tumor-infiltrating lymphocytes

3. Dendritic cells

4. T cells with chimeric receptors

5. T-cell receptor-activated cells

In cell therapy, live cells from healthy organs, fetuses or embryo of animals are injected into the patients. The injected cells find their way to weak or damaged organs of the same type and stimulate the body's own healing process.

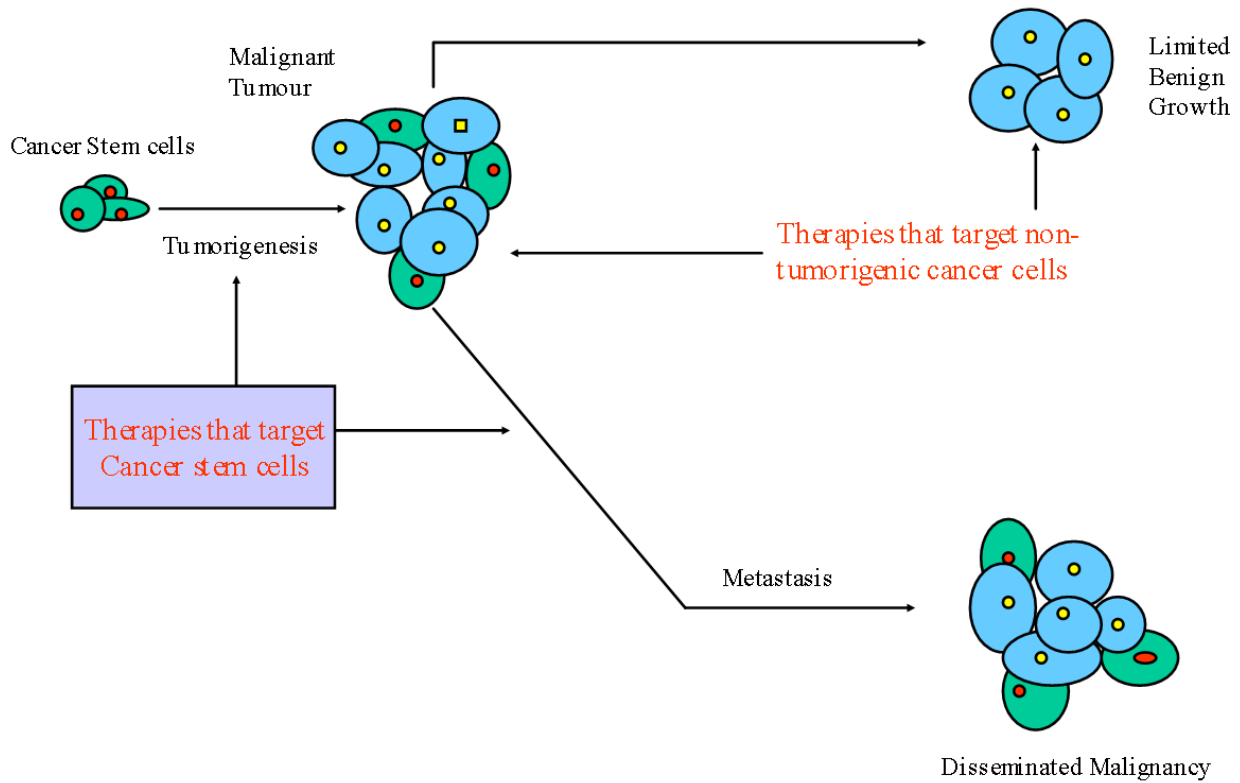

**Fig.16: Cellular therapy, mechanism of action**

[19]In cancer stem cell therapy, stem cells are derived from the inner cell mass of the blastocyte for embryonic stem cells, primary germinal layers of embryo for germinal stem cells and hematopoietic, neural, gastrointestinal and mesenchymal tissues for adult stem cells. In this therapy cancer cells are removed from stem cells. This technique has particular implications in bone tumors as reconstruction of bone following chemotherapy and surgery is always a serious problem.

The concept of stem cell as the delivery vehicles came in to the existence as the tumors send out chemoattractants such as the vascular endothelial growth factor (VEGF) to recruit

MSC to form the supporting stroma of the tumor, and pericytes for angiogenesis. It may be possible to deliver immune-activating cytokines and other secreted proteins to the brain and breast tumors though the stem cells. The major restriction in this therapy is the life of stem cells. Most of ASC do not possess sufficient telomerase activity and thus cannot prevent loss of telomerase. At each division, the telomeres shorten and the replication slows down (aging) and at the end, cells cease to divide (crisis phase). Thus, we may not be able to obtain enough adult stem cells to perform our clinical task.

[20]Dendritic cell therapy is also gaining strength in the field of cellular therapy. Dendritic cells are the immune cells which help in the recognition, processing and presentation of foreign antigens to T-cells in the effector arm of the immune system. In dendritic cell therapy, blood cells are harvested from the patients and are processed in the laboratory to produce dendritic cells which are then given back to the patient to allow massive dendritic cell participation in optimally activating the immune system.

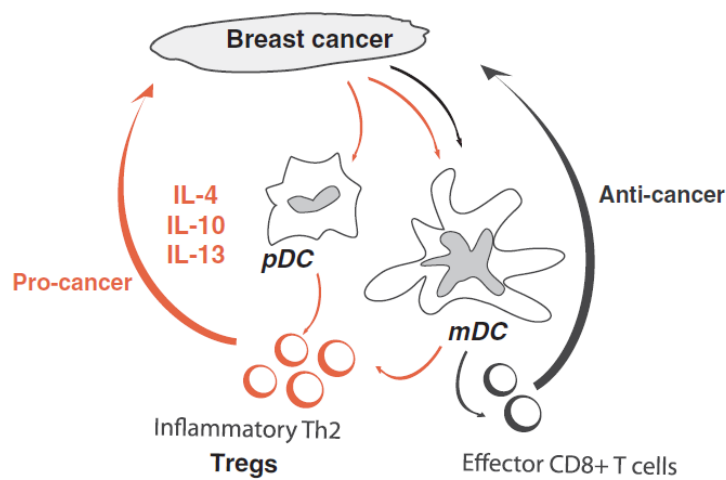

**Fig.17: Dendritic cells (DCs) in tumor environment**

Cancer cells attract immature DC possibly through chemokines such as MIP3 alpha and / or SDF-1. The DC can then be either blocked or skewed in their maturation, for example by VEGF, leading to induction of polarized CD4+ T cells that promote the expansion of cancer cells (pro-cancer) at the expense of CD8+ T cells that can cause tumor regression (anti-cancer). A compelling strategy would be to rewire their molecular pathways from pro-cancer DCs into anticancer-cancer DCs for example with antibodies or DC activators[20].

*Gene Therapy*[21]*:*

Genes are the blue print of heredity. They are responsible for traits and characteristics. Genes are located in chromosomes and made of deoxyribonucleic acid. Altering a person's genetic material can prevent diseases. Gene therapy is an experimental treatment in which genetic material is introduced in the patient to fight against the disease. This therapy is presently studied in clinical trials for various types of cancer, but it is not yet available outside the clinical trial.

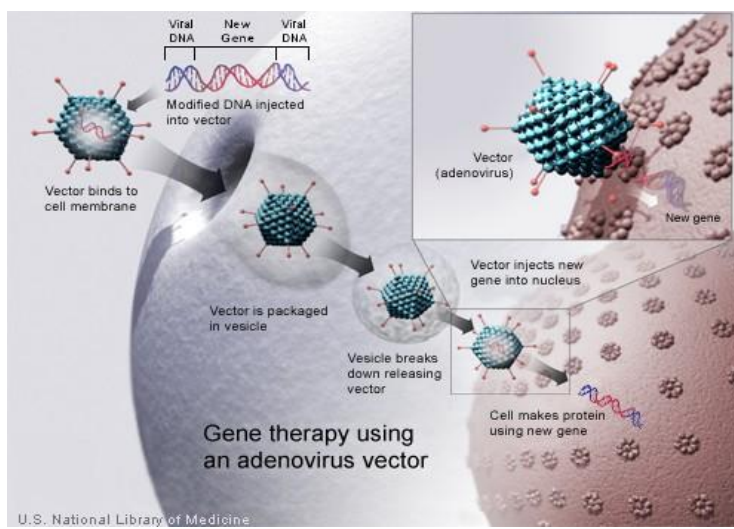

Fig.18: Gene Therapy, Targeted gene therapy: A treatment to remove cancer roots Cancer is molecular network disease

Researchers are studying many different ways to treat cancer with gene therapy. One of the approaches involves replacing the missing or altered gene with the healthy genes. Due to some missing or altered genes cancer may occur so it is a nifty idea to replace them with the "working" copies of those genes.

Another study involves stimulating the body's natural ability to attack cancer cells. In this approach researchers take a blood sample from patients and insert genes that will cause each cell to produce T-cell receptor (TCR). The genes are transferred into the patient's white blood cells (called T lymphocytes) and are then given back to the patient. The TCRs then recognize and attach to certain molecules found on the surface of the tumor cells. Finally, the TCRs activate the white blood cells to attack and kill the tumor cells.

Scientists are also investigating the insertion of genes to the cancer cells to make them sensitive to chemotherapy and radiotherapy. Healthy stem cells forming blood cells are removed, and the gene that makes these cells more resistant to the side effects of high doses of anti cancer drugs are inserted and finally injected back to the patients.

In another approach "suicide genes" are introduced in to the patient's cancer cells. A pro drug is then given to the patient, and the pro drug gets activated in cancer cells with the suicide genes ultimately leading to the destruction of the cancer cells.

Vectors play a vital role in this study as genes can not be directly inserted in the person's cell. Viruses are the common choice of vector in this case. They have the unique ability to recognize certain cells and insert genetic material into them. It involves removing cells from patient's blood or bone marrow and growing them in the laboratory. The cells are exposed to the virus that is carrying the desired gene. The virus enters the cells and inserts the desired gene into the cells. These cells then grow in the lab environment and are then injected back into the patient's vein. This is also called ex vivo gene therapy as the cells are grown outside the body.

*Controlled Amino Acid Therapy[22]*:

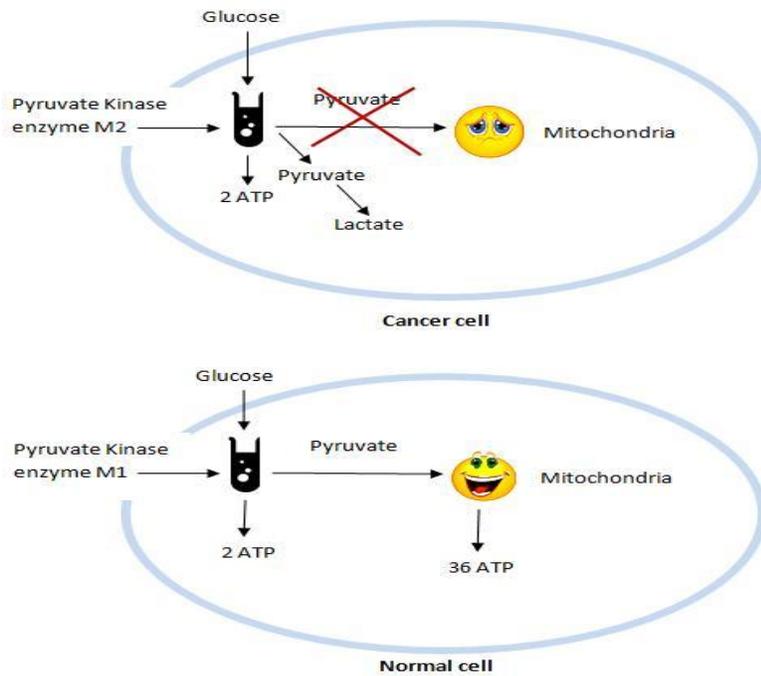

**Fig.19: Controlled amino acid therapy, metabolism pathway in cancerous cell vs. normal cell**

CAAT is a dietetic approach to cancer therapy. In cancer cells the citric acid cycle (Kreb's cycle) is inhibited. They claim their energy requirement mainly from glycolysis. So, if the glycolysis is specially inhibited, cancer cells, but not the normal cells, can be starved to death. The objective of CAAT is to alter or impair the development of cancer cells by interfering with the five basic requirements of cell formation (structure, energy, blood vessels, growth hormones and functions). This is accomplished by controlling the intake of the 20 different amino acids that the cancer cell requires for formation, growth and function. It impairs the creation of certain amino acids that are essential to manufacture DNA in cancer cells. Without structure, the cancer cell cannot develop. It increases the daily intake of certain amino acids, and along with its individualized formulation manipulates the two energy systems. The differences between how cells derive their energy explains why the functionality of normal cells will not be affected, and why the shutting off of the energy supply to cancer cells, by inhibiting glycolysis, results in starving them to death. Also, it inhibits production of enzymes and hormones that are essential to their growth, reproduction and metastasis.

# Methodology

**Specific Aims**

The thesis focused on various regimens of cancer biotherapy:

1. Building database in MySQL and UNIX server **(MySQL)**

2. Build a website for public access **(HTML)**

3. Link the database to the website: database driven website **(HTTP and PHP)**

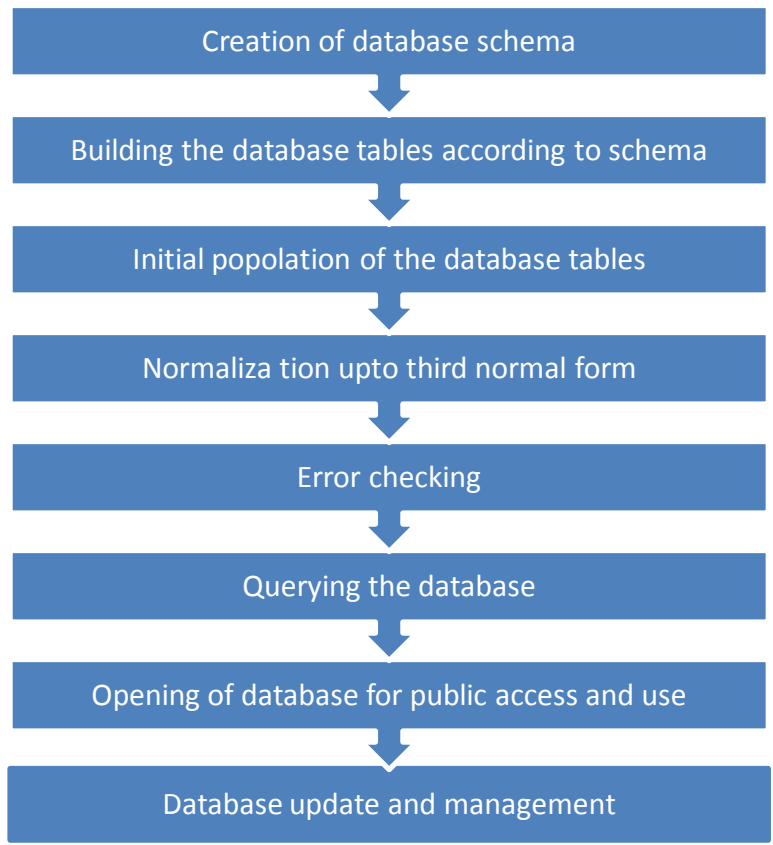

**Specific Aim 1: Building database in MySQL**

The database will have various regimens of cancer biotherapy which includes:

Protein Kinase Inhibitors, 2.) Ras Pathway Inhibitors, 3.) Cell-Cycle Active Agents, 4.) MAbs (monoclonal antibodies), 5.) ADEPT (Antibody-Directed Enzyme Pro-Drug Therapy), 6.) Cytokines (interferons, interleukins, etc.), 7.) Anti-Angiogenesis Agents, 8.) Cancer Vaccines (peptides, proteins, DNA), 9.) Cell-based Immunotherapeutics, 10.) Gene Therapy, 11.) Hematopoietic Growth Factors, and 12.) Retinoids 13.) CAAT.

For each biotherapy regimen, we will extract the following attributes in populating the database: (a.) cancer type, (b.) gene/s and gene product/s involved, (c.) gene sequence (GenBank ID), (d.) organs affected, (e.) available chemo treatment, (f.) reference papers, (g.) clinical phase/stage, (h.) survival rate (chemo. vs. biother.), (i.) clinical test center locations, (j.) cost, (k.) patient blog, (l.) researcher blog (k) Future work.

The database is created in MySQL server. The database consists of 9 tables: tbl_cost, tbl_cancer_type, tbl_organs, tbl_clinical_trial, tbl_chemo, tbl_survival_rate, tbl_location, tbl_sequence.

All the tables have a primary key and other attributes:

```
create table cost(

CostID int,

TreatmentID int,

TreatmentCost decimal(10,2),

CONSTRAINT cost_CostID_pk PRIMARY KEY(CostID)

)
```

The main table, tbl_biotherapy_agent is connected by other tables by foreign key:

```
CONSTRAINT cost_TreatmentId_fk FOREIGN KEY (TreatmentId) REFERENCES treatment(TreatmentID)
```

The tables are populated from the information provided by research papers and work in the field of cancer biotherapy.

Database:

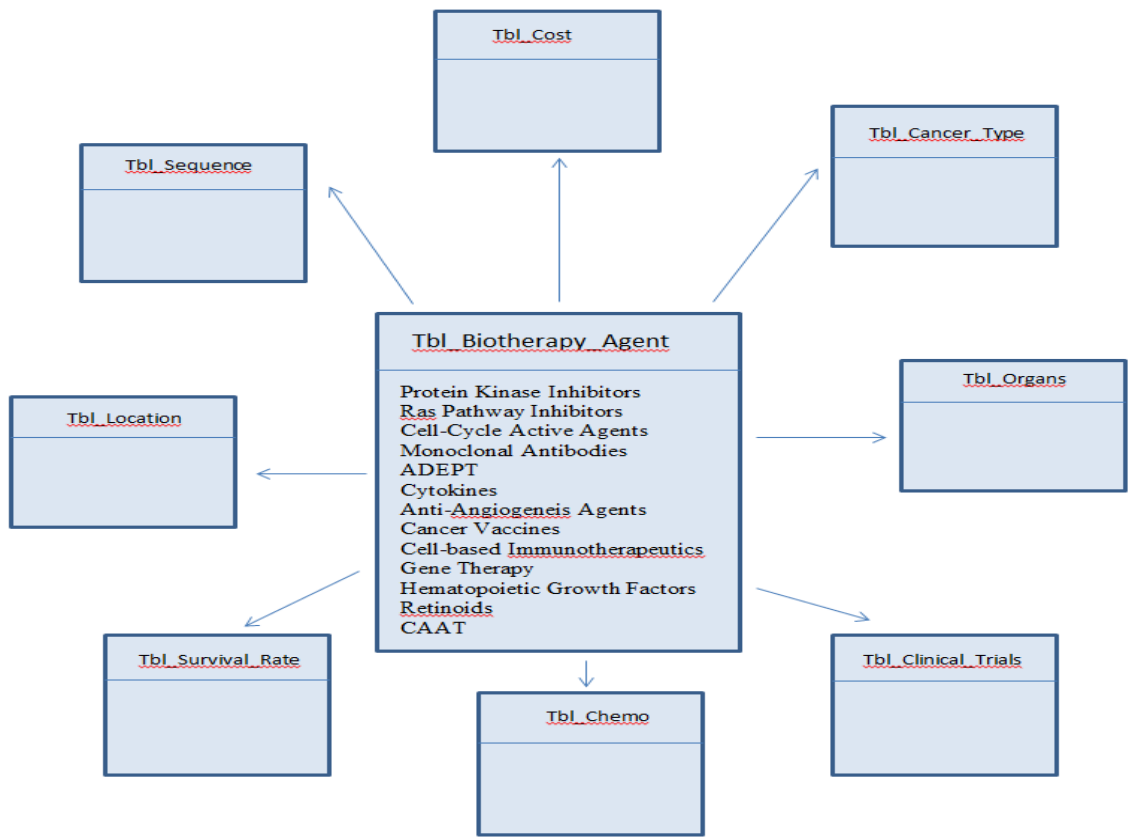

**Fig.20: Cancer Biotherapy Database**

**Fig.21: Schema of the database**

Schema is the structural explanation of the database. It shows the relation between the tables in the database.

**Step 2: Build a website for public access (html)** [23]

Website is a set of web pages (like texts, images, videos etc.) hosted by web servers via a network such as the Internet or a private local area network through an Internet address known as a Uniform Resource Locator (URL). All publically available websites constitute World Wide Web (www).

The website is built in HTML and PHP to make it dynamic and is available for public access. HTML is Hyper Text Markup Language which uses markup tags to describe web pages. PHP is a general-purpose server-side scripting language originally designed for Web development to produce dynamic Web pages. The website will be database driven in addition to information about the biotherapy regimens as discussed above, the website will be connected to other biotherapeutically relevant web servers. The website will also contain a news page for the updated information in the field of cancer and biotherapy.

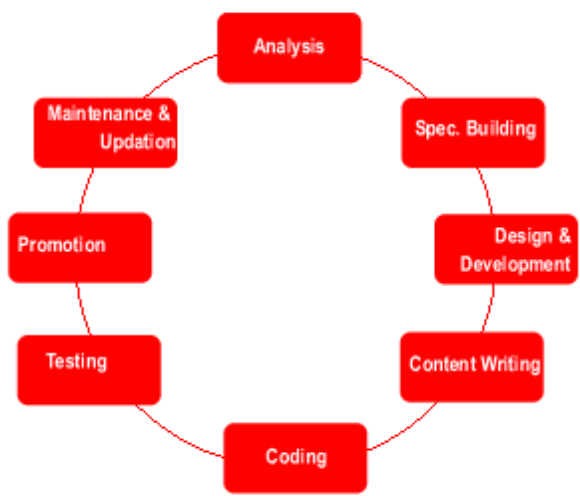

**Fig.22: Website Building**

For building a website it is necessary to analyze the needs of the website and its contribution to the process and project. The analysis helps in the realization of the need for the website. It is necessary to know how the development of the website will connect various existing systems and information. For building our website it is necessary to know its function as an information platform. The website is needed to give information about various cancer biotherapy treatments. It is also providing a discussion platform for researchers. The website will also hold experiences of patients and survivors and will provide information to layman about present treatments.

The next step in the process is specification building. Preliminary specifications are drawn up by covering up each and every element of the requirement. It helps in sketching the outline of the website as per the requirement. Our website contains a banner with the name of the database, navigation bar, sidebars, main body and footer for copyrights.

Designing and development is the next step in building the website. The designing of the website is done in CSS and PHP. It involves the template selection, imaging and prototypes. The content consists of explanation and information about the database and different regimens of the treatment. It involves the news updates, discussion blogs and frequently asked questions. The treatment regimen gives information of different biotherapies, its recent advancements and availability.

With all the planning finally the coding is done using html, PHP, CSS and java script. The database is built in MySQL. The database was connected with the website using java script. WebPages is coded and interlinked with html and PHP. The styling of the website is done in CSS.

The testing is done by populating the database and checking its reflection in the website. The navigation bar was tested for other web pages and drop down menu. The side bars were tested for web pages and news updates. The drop down menu directs to the various treatment regimens connected with the database. The blogs let the people share their experience and queries and the frequently asked page gives additional information about cancer and the website.

This database driven website will be maintained manually and updated every 2 months. The promotion is future work and would be done by linking the website to cancer.gov.

**Step3: Link the database to the website: database driven website (http & PHP)**

The Hypertext Transfer Protocol (HTTP) is an application protocol for distributed, collaborative, hypermedia information systems. The database will be connected to the website to provide information about the various cancer biotherapy regimens. The website connected with the database is known as a database driven website and is generally done with PHP. To have a database connected to the website is always beneficial as it makes it easier to introduce any further updates to the website by just modifying the data once. The database is connected to the website in java scripting language. In the navigation bar of the website, the treatment option has a drop down menu. The drop down menu holds names of the treatment regimens. Clicking on any of the treatment fetches information about that treatment from the database and reflects it as a webpage in the website.

## Results and Discussion

## Database:

The database was named Biotreatment. It was populated in PhPMyAdmin and queried with MySQL. As per the schema, the main table, treatment, was connected with all other information tables using a foreign key.

```
mysql> select * from treatment;
+-------------+-------------------------------+
| TreatmentID | TreatmentMethod               |
+-------------+-------------------------------+
|           1 | Protein Kinase Inhibitors     |
|           2 | RAS Pathway Inhibitors        |
|           3 | Cell-Cycle Active Agents      |
|           4 | Monoclonal Antibodies         |
|           5 | ADEPT                         |
|           6 | Cytokines                     |
|           7 | Anti-Angiogenesis Agents      |
|           8 | Cancer Vaccines               |
|           9 | Cell-based Immunotherapeutics |
|          10 | Gene Therapy                  |
|          11 | Hematopoetic Growth Factors   |
|          12 | Retinoids                     |
|          13 | CAAT                          |
+-------------+-------------------------------+
```

**Fig: Cancer Biotherapy Database**

## Website:

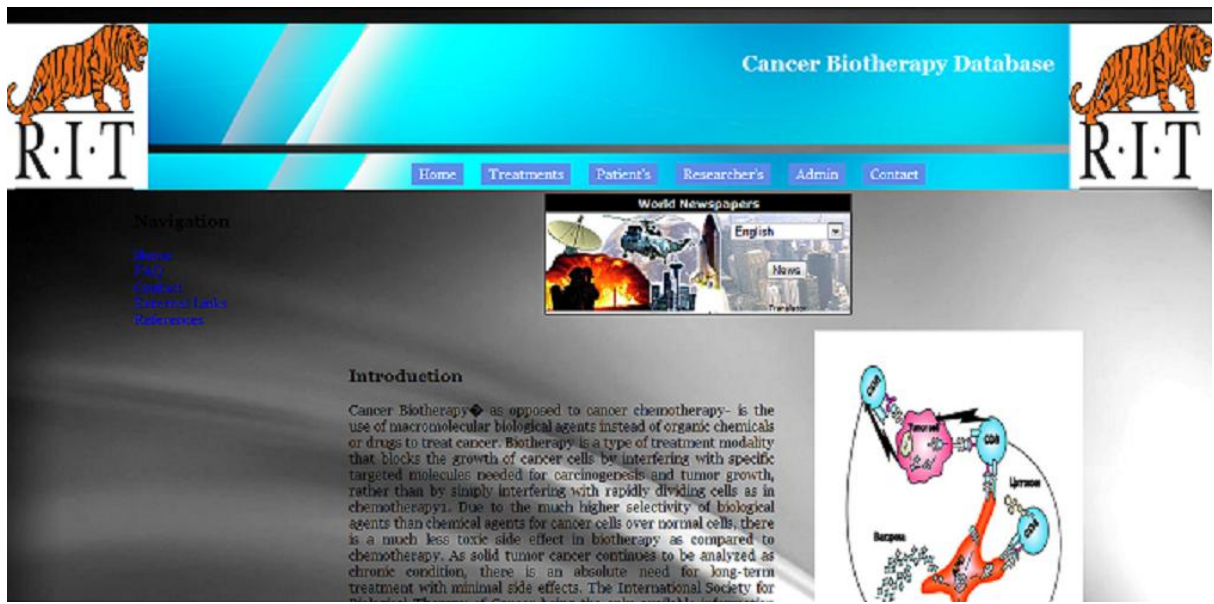

**Fig.23: Website**

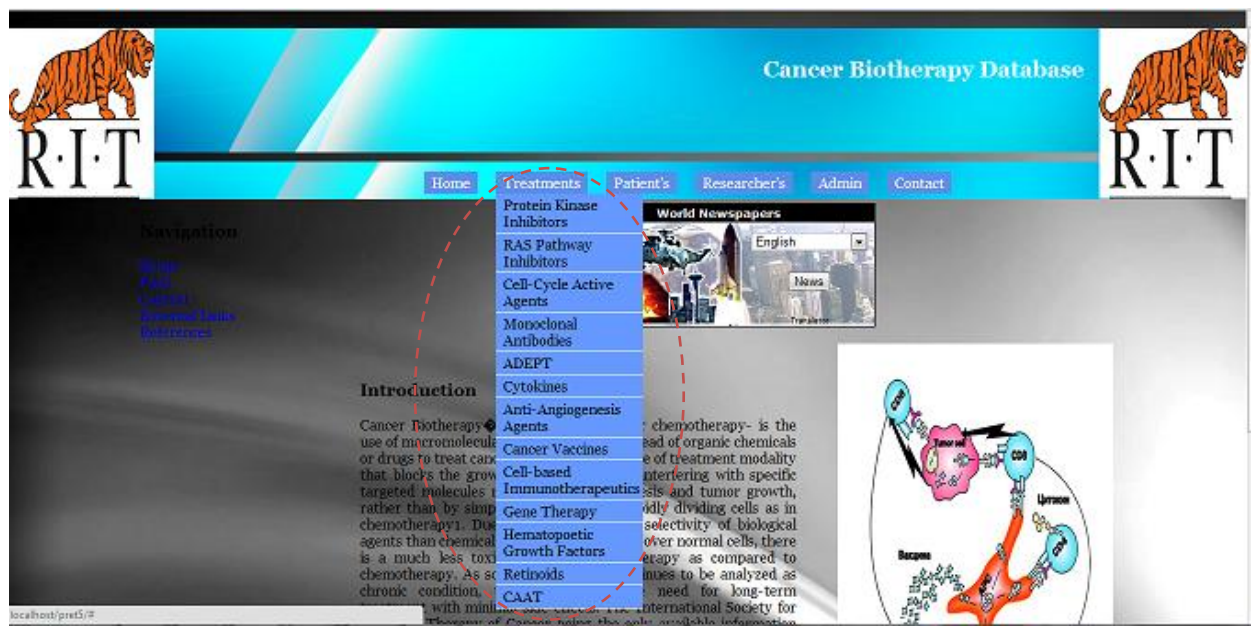

**Fig.24: Navigation bar with treatment database connection**

The website has a banner with title "Cancer Biotherapy Database" flanked by RIT logo. Below the header is the navigation bar with tabs: Home, Patient, Researcher, and Contact. The Home tab brings to the introductory page which gives the general idea about biotherapy. It reflects the code from index.php for the website. Treatment is the next tab which holds a drop down menu of different regimen of Biotherapy. The treatment tab is linked to the database and fetches information about each type of treatment directly from the database. The Patient tab is a blog for patient's to share their experiences. It is available for all users to read and write. The Researchers tab is a discussion platform for the researcher to discuss their work and give information about latest advancements in the field of cancer biotherapy. Contact tab holds information about the developer of the project. Below the navigation bar is the NEWS updates which provides information about the latest NEWS updates in various languages. The Sidebars holds: Home, External Links and References. The External Links give information about other proteomics, metabolomics, glycomics and lipidomics databases. References include the addresses which were referred in a building project and achieving informations about the treatments.

**Below is information about the different biotherapy regimens found from various sources:**

**Cytokines**[24, 25]**:**

Some of the cytokines have direct effects on tumors. They stimulate the growth of cells in immune system. Cytokines involved in cancer therapy are alpha/beta interferons, interleukin-2 & 12 and gamma interferon.

Interferons interact with more than 20 chemotherapeutic agents. They increase or decrease in the metabolism of chemotherapeutic agents. They are administered by subcutaneous injecting daily in low doses for CML or three times weekly in higher doses for solid tumors. Its doses range from 3Mu/day to 10Mu/tiw to 20 Mu daily for 1 month. Some of the side effects of IFN are fever, chills following injection, loss of appetite, headache, myalgias, arthralgia and cumulative fatigue.

IL-2 activity was tested in vitro in animal models for methylcholanthrine induced sarcomas, colorectal cancer, melanoma, leukemia and lymphoma. It is administered in high doses in short course of 2 weeks. Subcutaneous injections can be taken daily x 5 or tiw for prolonged course of 6 months. It can be administered only in cases of no active brain metastases, no medical contraindications and metastatic RCC or melanoma. Some of the side effects of the treatment involve hypotension, tachycardia, fever, chills, nausea, erythematous rash and oliguria.

Il-2 had produced durable complete response in stage IV melanoma and renal cell cancer. This treatment is particularly effective in solid tumor except testicular cancer.

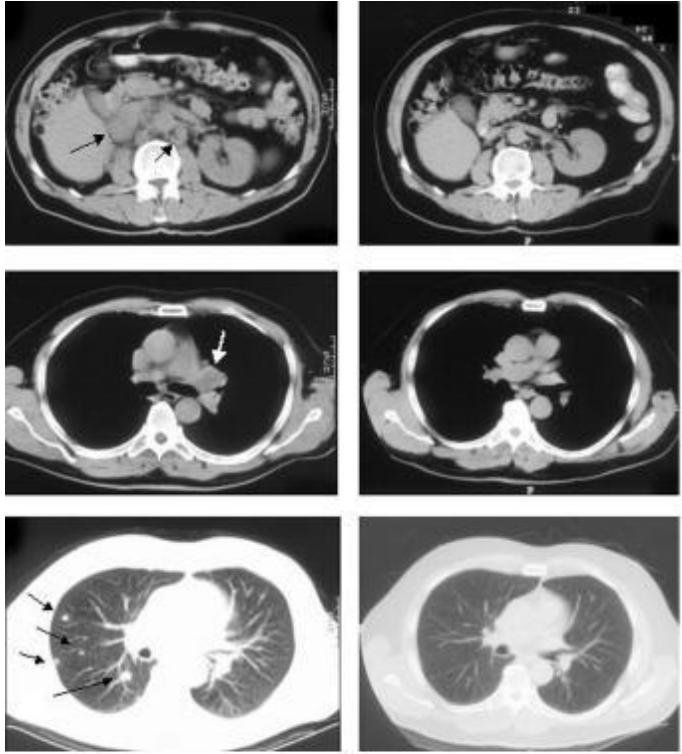

Figure 5. A patient with baseline lung metastasis and lymph node metastases in retroperitoneum and left hilus (left lane, arrows). The patient received IL-2 and histamine and achieved complete response. The patient is alive with no recurrence after 43 months of follow-up.

**Fig.25: IL-2 based immunotherapy in renal cell carcinoma (RCC)**
http://dadlnet.dk/dmb/DMB_2007/0407/04-07-disputatser/DMB3947.pdf

Other cytokines studied are IL-4 for low grade B cell malignancies, IL-6 for solid tumors, IL-12 for melanoma, RCC, CTCL but its toxicity stopped further development and IFN-y (not active in-vivo).

**Retinoids**[26, 27]**:**

Retinoic acid interferes with the growth and development of cells. It is related to vitamin A. Retinoic acid is administered by mouth. It comes in capsule form and should be swallowed whole with food or milk. It should be given at regular times to keep a steady level in the bloodstream. The side effects during the treatment is dry skin, swollen lips, skin rash, fatigue and eye irritation.

Retinoid most extensively studied in cancer therapy is synthetic amide of retinoic acid *N*-(4-hydroxyphenyl) retinamide (4-HPR), or fenretinide. Fenretinide has been found to exert significant chemo preventive activity in a large variety of *in vitro* and *in vivo* systems. It inhibits the proliferation of breast cancer cells which don't express RAR.

Both fenretinide and its principal metabolite 4-methoxy-phenylretinamide (4-MPR) selectively accumulate in the human breast ([Mehta *et al.* 1991](#)) rendering this agent an attractive candidate for breast cancer chemoprevention. A phase I dose ranging study was completed, and a 200 mg daily dose was selected as the safest one ([Costa *et al.* 1989](#)). Studies of the mechanisms responsible for retinol reduction have indicated that fenretinide has a high binding affinity to retinal-binding protein (RBP), thus interfering with the RBP–retinol–transthyretin complex formation and the secretion of retinol from the liver ([Berni & Formelli 1992](#)).

**Monoclonal Antibody[7]:**

Monoclonal antibodies can be used in various ways in the cancer therapy. They act through ADCC and CDC. Alternatively, they can conjugate with the toxin, cytotoxic agent or radioisotope to act against the cancer cell antigen. With the conjugation a toxin is bound to the antibody, which eventually attaches to the antigen on the cancer cell. The antibody conjugate is absorbed into the cell itself, resulting in cell death. By conjugating with a radioisotope like iodide-131 can directly infuse the cancer cells with radiotherapy. Also, we can attach chemotherapeutic agents which can be directly taken in to the targeted malignant cell.

In monoclonal therapy one of the most common targets is CD20, found in B cell malignancies. The CD52 antigen is targeted by alemtuzumab in treatment of chronic lymphocytic leukemia, CD22 is targeted in many cancer cases and has recently shown efficacy in hairy cell leukemia. Two newly introduced monoclonal antibodies targeting CD20 are tositumomab and ibritumomab.

**Table 1. Monoclonal Antibodies for Treatment of B cell Malignancies**

| Antigen | Antibody | Type | Investigational Status |
|---|---|---|---|
| CD20 | Rituximab (Rituxan) | Chimeric | FDA approved |
| | Tositumomab (Bexxar) | $^{131}$I-Murine | Submitted |
| | Ibritumomab (Zevalin) | $^{90}$Y-Murine | Submitted |
| CD52 | Alemtuzumab (Campath) | Humanized | FDA approved |
| CD22 | Epratuzumab (Lymphocide) | Humanized | Phase II/III |

**Fig. 26: Monoclonal Antibody Treatment**

The very first monoclonal antibody approved by FDA was Rituximab. Rituximab is a chimeric unconjugated monoclonal antibody directed at the CD20 antigen. It is indicated for the treatment of low-grade lymphomas. Results of studies using Rituximab as first-line treatment of low-grade non-Hodgkin lymphoma have been encouraging.[31-33] Patients who had not received any prior therapy were treated with Rituximab 375 mg/m² on a weekly basis for 4 weeks and then re-evaluated 2 weeks post-therapy. The patients who achieved a complete or partial response, or who had stable disease received Rituximab maintenance therapy (weekly for 4 weeks every 6 months). Patients who showed evidence of progression were taken off maintenance therapy. During re-evaluation at 6 weeks 54% of the patients showed a response to the treatment and 36% had stable disease or minor response.

Also, Rituximab has been studied in combination with chemotherapy for patients with intermediate grade or diffuse large cell non-Hodgkin lymphoma. The therapy is called CHOP (cyclophosphamide, doxorubicin, vincristine, and prednisone). In a study with 33 patients having newly diagnosed large cell lymphoma, the response was 94%. Out of which 61% (20 patients) achieved complete response, 33% (11 patients) received partial response. These results were confirmed in phase III trial of CHOP and Rituximab in elderly patients conducted by the French Lymphoma Cooperative Group (GELA).[35]

Monoclonal antibodies in studies of solid tumor is edrecolomab and trastuzumab. Endrecolomab targets 17-1A antigen seen in colon and rectal cancer and has been approved for use in Europe for these indications.[36, 37] Its actions are mediated through ADCC and CDC. In a study with 189 with stage II colorectal cancer, treatment with edrecolomab reduced the mortality risk by 32%. Trastuzumab is one of the most commonly used monoclonal antibody for the treatment of solid tumors. It targets HER-2/neu antigen which is found in 35% of breast cancers. Trastuzumab down regulates HER-2 receptor expression. In phase I and II trials of patients with metastatic breast cancer, treatment with a combination of trastuzumab and cisplastin resulted in prolongation of survival and higher response rates. In combination with the chemotherapy prolonged the survival rate by 7.4 months.

**Cancer Vaccines**[12, 13, 14, 15]**:**

Cancer vaccines are fluid administered by injecting under the skin. Their dose depends on the type of cancer and type of vaccine. Most treatments in research trials focus on advanced cancers of the prostate gland, breast, pancreas, colon and rectum, lung, skin, kidney, ovary, bladder, cervix, leukemia and lymphoma.

**Table 3. Tumor Cell Vaccines**

| Vaccine | Immune Response | Clinical Response | Reference |
|---|---|---|---|
| Melanoma transduced with Ad-GM-CSF | 5/9 + CTL activity | 1/9 minor response | Cancer Immunol Immunother. 2001;50:373-381. |
| Allogeneic panc. tumor secreting GM-CSF | 3/14 DTH+ | 3 patients with DFS >25mo | J Clin Oncol. 2001;19:145-156. |
| CancerVax (melanoma) | 82% + complmnt dependent cytotox (CDC) | Median survival 54 mo, if deltaCDC ≥ 10% | Ann Surg Oncol. 1998;5:595-602. |
| Autologous colon CA + BCG | Increased DTH in all patients | No overall survival benefit | Vaccine 2001;19:2576-2582. |
| Autologous GBM+ Newcastle Virus | DTH increased from 1.67 to 4.05 cm² | Median survival was 46 weeks | J Neurooncol. 2001;53:39-46. |

**Fig.27: Cancer Vaccine Clinical trials**

The table illustrates results of research on tumor vaccines. These studies employed several strategies. One of the most popular is to transducer with a vector containing GM-CSF, so that the tumor secretes GM-CSF and sets up an inflammatory response.[68,69] In animal studies, this strategy has been the most promising in inducing a protective immune response. Newcastle virus is used to infect the cells in another study[72] to some effect, as shown by median survival of 46 weeks. Use of Bacille Calmette-Guerin (BCG) as an inflammation inducing adjuvant along with autologous colorectal cancer cells[71] showed increased DTH but no survival benefit. In fact, most of these tumor cell approaches show an immune response, but again limited clinical response. Nonetheless, CancerVax[70], which has been heavily tested in melanoma, has been suggested in nonrandomized studies to provide a survival benefit.

**ADEPT**[29, 30]**:**

Antibody directed enzyme prodrug therapy is the bio-chemotherapy and falls under the regimen of targeted therapy. ADEPT is only being used in clinical trials. The trials aim to find out whether ADEPT may be useful as a new type of treatment for bowel cancer. It uses monoclonal antibody to carry enzyme directly to the cancer cells.

ADEPT is a colorless fluid. It is given by drip or infusion through a small tube inserted in vein. The monoclonal antibody and the prodrug are usually given in two separate doses on the same day. Some of the pre-clinical studies were done using in-vivo models in the labs of UCL Cancer Institute by Dr. Surinder K Sharma. *In vitro* and *in vivo* studies are in progress to select suitable candidates for enhanced ADEPT.

The group has critical responsibilities in relation to clinical trials. They develop improved techniques for laboratory measurement of parameters that are important to the care, safety and management of patients. One essential feature of the ADEPT clinical trials has been the need to monitor the patient's blood enzyme levels to ensure that prodrug is given only when it is safe to do so. These measurements are performed by the group in real time.

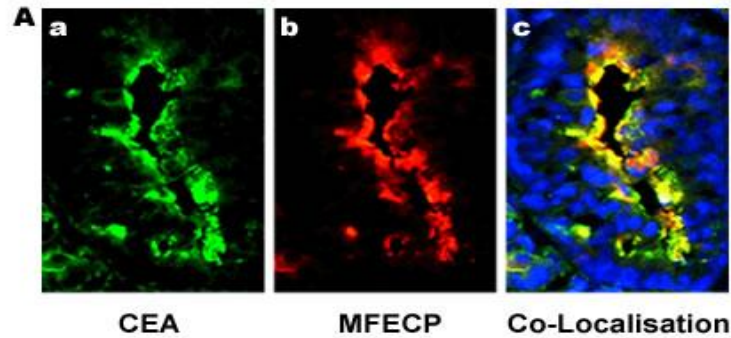
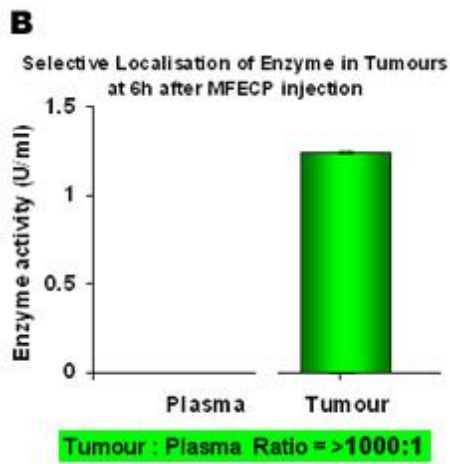
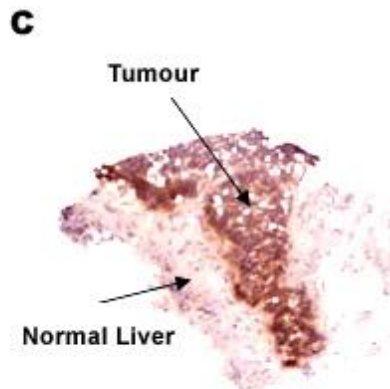

**Fluorescence images of human colon carcinoma cells showing selective localization of antibody-enzyme fusion protein (MFECP) in-vivo.**
**Panel A**: a) Expression of CEA, b) Localization of i.v. given MFECP fusion protein, c) Co-localization of CEA and MFECP (yellow) and DAPI (blue) showing tumor cell nuclei (x 400 magnification). [Sharma et al, Clin Cancer Res., 2005]
**Panel B**: Selective retention of enzyme activity in tumor and rapid clearance of enzyme activity from blood with MFECP in-vivo.
**Panel C**. A liver biopsy specimen from an ADEPT patient showing localization of antibody-enzyme (MFECP) in metastatic tumor cells. [Mayer et al, Clin Cancer Res., 2006]

## Fig.28: ADEPT Trials

http://www.ucl.ac.uk/cancer/oncology/adept

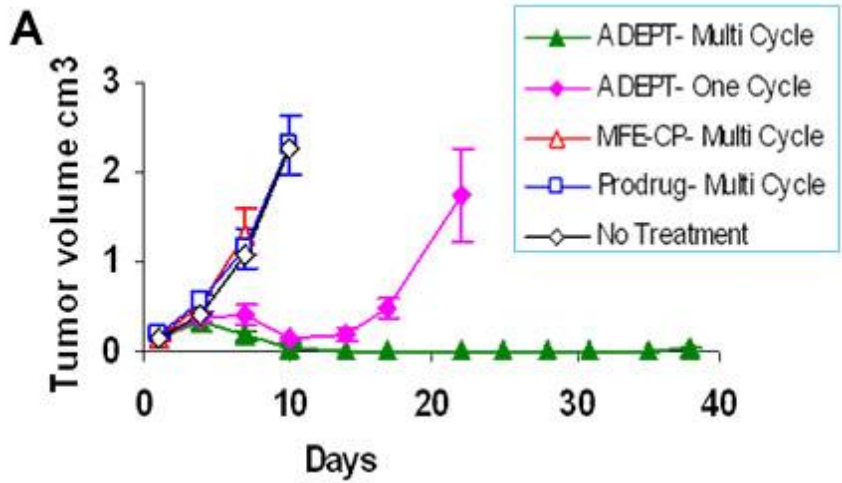

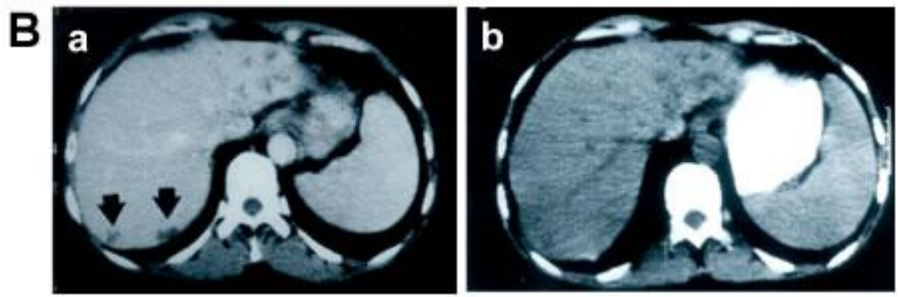

**Panel A**: Multiple cycles of ADEPT result in tumor regression in in-vivo.
**Panel B**: **a)** Metastatic tumor deposits of colorectal cancer in Liver (arrows) before treatment and **b)** tumor regression after ADEPT treatment [Napier et al, Clin Cancer Res., 2000]

### Fig.29: ADEPT treatment responses

They are currently exploring new combination therapies to enhance tumor cell death without affecting the normal tissues. These include studies of new prodrugs, inhibitors of DNA repair, anti-vascular, anti-angiogenic agents as well as and complimentary antibodies, which may synergise with ADEPT for the treatment of a variety of different tumor types[31, 15].

**Anti-Angiogenesis Agents**[16, 17]:

Anti-angiogenesis research began more than 35 years ago with the work of the late Judah Folkman, MD. This therapy doesn't affect the same parts of the body as of chemotherapy. While chemotherapy causes the cancer cells to shrink or disappear, angiogenesis inhibitors stops them from growing any larger. These two treatments can be combined and given. There are two angiogenesis inhibitors studied for the cancer therapy: lenalidomide and thalidomide.

Lenalidomide is immunomodulatory drug. It affects the way immune system works and also blocks the development of new blood vessels. It can be only used to treat people who had already been under one of the treatments already. It is generally accompanied with steroid dexamethasone and sometimes with the chemotherapeutic drug cyclophosphamide. Lenalidomide is available as 5mg, 10mg, 15mg and 25mg capsules. The capsules are swallowed whole with plenty of water. Possible side effects during treatment are neutrophia, bruising, increased risk of blood clots, rashes, fatigue and nausea.

Thalidomide is mainly used in the treatment of myeloma. It is accompanied by melphalan and prednisolone. They are available as 50 mg capsules and should be swallowed with water an hour after food. They can also be given in combination with chemotherapy drugs. Side effects during treatment with thalidomide are nausea, vomiting, temporary reduction in blood cells, numbness, headache, dizziness, skin changes and ankle swelling.

**Gene Therapy**[21]:

Gene therapy involves the correction of genetic defects. The gene therapy approach includes transduction of drug resistance genes to protect bone marrow, sensitization of tumor cells for destruction by the therapeutic agents, replacements of tumor suppressor genes and inactivating oncogenes. Viruses used as vectors in the gene therapy are retroviruses, adenoviruses, adeno-associated viruses, lentiviruses, poxviruses, and herpes viruses. All these viruses differ their way of transferring genes to the cells they recognize and so they are picked based on the requirement of the study. Investigators alter the viruses to make them safe for human and to enhance their ability to deliver specific genes to the patient's cells.

In a clinical trial of lung cancer treatment, gene therapy was used in combination with radiotherapy in 19 different patients. Patients were injected Ad-p53 with intramural needle on days 1, 18 and 32 of the treatment accompanied by the radiation therapy. 17 out of 19 patients made it through the entire therapy.

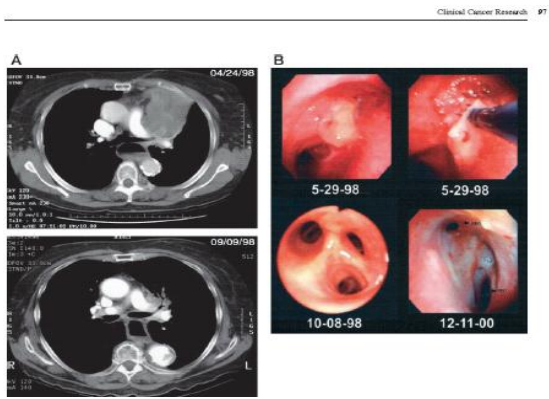

**Fig.30: A, patient no. 2: left upper lobe tumor unable to undergo surgery because of poor pulmonary function and cardiac disease. Patient received three injections of Ad-p53 (3 _ 1011 vp) via bronchoscope in combination with radiation therapy (60**

**Gy; A). Pathologic biopsy negative for viable tumor 3 months after completion of therapy (B). B, patient no. 3: right upper lobe tumor unable to be treated with surgery because of poor pulmonary function and ineligible for chemotherapy because of cardiac disease and obstructed bronchus (5/29/98). Patient was treated with three injections of Ad-p53 (3 _ 1011 vp) and radiation therapy (60 Gy) by bronchoscopy (5/29/98) with a CR 3 months after completion of therapy (10/8/98) and no pathologic evidence of tumor 29 months after therapy (12/11/00). 57% of the patients showed worsening growth stage of the cancer.**

Problems faced in the experiment were to ensure if the vectors have successfully inserted the genes in the desired target cells and to ensure that the translated gene is controlled by body's normal physiologic signals.

The focus of the building this database is to provide adequate information about biotherapy to the patients, their relatives and laymen. This database also proves informative to the researchers with all the information about the therapy and its cross linkage with other database. This is the first database which focuses on the biotherapy regimen for the cancer research and treatment. It also provides the discussion platform to the patients to share their experience and put forth their queries. The discussion platform also helps the researchers to exchange their ideas and gather more information about the research work going on in different parts of the world. At present the database is in its basic level so it needs more information to be updated. Also, the database needs a search function to make it easier for the user to look for a similar kind of information.

## Conclusion:

Biotherapy as the name suggests is the therapy which aids the body's own immune system to fight against cancer. It is a cancer cell specific treatment which suppresses the process that allows cancer growth, helps the immune system to identify and differentiates the cancer cells from normal cells and promotes the body's natural ability to repair or replace cells that have been damaged by cancer treatments. It is a wise idea to conjugate with other treatment modalities, such as surgery, radiation therapy and chemotherapy. A cumulative database is a boon to patients as well as researchers. It was the first database focusing on cancer biotherapy. In addition to the biological relevant information about the treatments, the database had patient's discussion blogs for better understandings of patients and for enabling them to talk about their experiences and researcher's blog to enable cancer biotherapy researchers to contact each other and suggest new ideas to other researchers.

# FIGURE REFERENCES

# Appendix

MySQL code for creating table in the database:

```
drop table if exists cost;
drop table if exists organ;
drop table if exists cancerType;
drop table if exists survival;
drop table if exists location;
drop table if exists trial;
drop table if exists chemo;
drop table if exists treatment;

CREATE TABLE treatment(
TreatmentID int,
TreatmentMethod varchar(40),
CONSTRAINT treatment_TreatmentID_pk PRIMARY KEY(TreatmentID));

create table cost(
CostID int,
TreatmentID int,
TreatmentCost decimal(10,2),
CONSTRAINT cost_CostID_pk PRIMARY KEY(CostID),
CONSTRAINT cost_TreatmentId_fk FOREIGN KEY (TreatmentId) REFERENCES
treatment(TreatmentID)
);

create table cancerType(
CancerTypeID int,
TreatmentID int,
CancerType varchar(40000),
CONSTRAINT cancerType_CancerTypeID_pk PRIMARY KEY(CancerTypeID),
FOREIGN KEY (TreatmentId) REFERENCES treatment(TreatmentID)
);

create table organ(
OrganID int,
TreatmentID int,
OrganType varchar(40),
constraint organ_OrganType_pk PRIMARY KEY(OrganID),
FOREIGN KEY (TreatmentId) REFERENCES treatment(TreatmentID)
);

create table trial(
trialID int,
TreatmentID int,
Trial_Phase varchar(40),
constraint trial_trailID_pk primary key(trialID),
FOREIGN KEY (TreatmentId) REFERENCES treatment(TreatmentID)
);
```

```
create table chemo(
ChemoID int,
TreatmentID int,
Chemo_Therapy varchar(400000000000),
constraint chemo_ChemoID_pk primary key(ChemoID),
FOREIGN KEY (TreatmentId) REFERENCES treatment(TreatmentID)
);

create table survival(
SurvivalID int,
TreatmentID int,
Survival_Rate varchar(40),
constraint survival_survivalID_pk primary key(SurvivalID),
FOREIGN KEY (TreatmentId) REFERENCES treatment(TreatmentID)
);

create table location(
LocationID int,
TreatmentID int,
TreatmentLocation varchar(40),
constraint location_LocationID_pk primary key(LocationID),
FOREIGN KEY (TreatmentId) REFERENCES treatment(TreatmentID)
);
```